\documentclass[onecolumn,draftclsnofoot,nofonttune,10pt]{IEEEtran}

\usepackage{amsmath}
\usepackage{amssymb}
\usepackage{amsfonts}
\usepackage{amsthm}

\usepackage[utf8]{inputenc}
\usepackage[T1]{fontenc}
\usepackage{txfonts}
\let\mathbb=\varmathbb

\usepackage{dsfont}

\usepackage[%
cal=cm,
frak=euler
]
{mathalfa}

\usepackage[dvipsnames,svgnames]{xcolor}
\colorlet{MyBlue}{DodgerBlue!40!Black}
\colorlet{MyGreen}{DarkGreen!80!Black}

\usepackage{subfigure}
\usepackage{tikz}
\usetikzlibrary{calc,patterns}

\usepackage{acronym}
\usepackage{balance}
\usepackage{latexsym}
\usepackage{paralist}
\usepackage{wasysym}
\usepackage{xspace}

\usepackage[numbers,sort&compress]{natbib}

\usepackage{hyperref}
\hypersetup{
draft=false,
colorlinks=true,
linktocpage=true,
pdfstartview=FitH,
breaklinks=true,
pdfpagemode=UseNone,
pageanchor=true,
pdfpagemode=UseOutlines,
plainpages=false,
bookmarksnumbered,
bookmarksopen=false,
bookmarksopenlevel=1,
hypertexnames=true,
pdfhighlight=/O,
urlcolor=Maroon,linkcolor=MyBlue,citecolor=MyGreen, 
pdftitle={},
pdfauthor={},
pdfsubject={},
pdfkeywords={},
pdfcreator={pdfLaTeX},
pdfproducer={LaTeX with hyperref}
}

\newcommand{\bA}{\mathbf{A}}

\newcommand{\bH}{\mathbf{H}}
\newcommand{\bI}{\mathbf{I}}

\newcommand{\bQ}{\mathbf{Q}}

\newcommand{\bU}{\mathbf{U}}
\newcommand{\bV}{\mathbf{V}}
\newcommand{\bW}{\mathbf{W}}
\newcommand{\bX}{\mathbf{X}}
\newcommand{\bY}{\mathbf{Y}}
\newcommand{\bZ}{\mathbf{Z}}

\newcommand{\bp}{\mathbf{p}}

\newcommand{\bw}{\mathbf{w}}
\newcommand{\bx}{\mathbf{x}}
\newcommand{\by}{\mathbf{y}}
\newcommand{\bz}{\mathbf{z}}

\newcommand{\C}{\mathbb{C}}
\newcommand{\R}{\mathbb{R}}

\DeclareMathOperator*{\argmax}{arg\,max}
\DeclareMathOperator*{\argmin}{arg\,min}

\DeclareMathOperator{\bigoh}{\mathcal O}

\DeclareMathOperator{\diag}{diag}

\DeclareMathOperator{\ex}{\mathbb{E}}

\DeclareMathOperator{\hess}{Hess}

\DeclareMathOperator{\prob}{\mathbb{P}}

\DeclareMathOperator{\tr}{tr}

\newcommand{\eps}{\varepsilon}

\newcommand{\mgeq}{\succcurlyeq}

\newcommand{\mleq}{\preccurlyeq}

\newcommand{\simplex}{\Delta}
\newcommand{\wilde}{\widetilde}

\newcommand{\smallabs}[1]{\lvert #1 \rvert}
\newcommand{\norm}[1]{\left\| #1 \right\|}
\newcommand{\smallnorm}[1]{\| #1 \|}

\newcommand{\given}{\:\vert\:}

\newcommand{\txs}{\textstyle}

\newcommand{\insum}{\sum\nolimits}

\usepackage[textwidth=30mm]{todonotes}
\usepackage{soul}
\setstcolor{red}
\sethlcolor{SkyBlue}

\usepackage{algorithm2e_extension}
\usepackage{algorithmic,algorithm}
\SetKwBlock{Repeat}{Repeat}{}

\newtheorem{theorem}{Theorem}

\newtheorem*{corollary*}{Corollary}

\newtheorem{proposition}{Proposition}

\theoremstyle{definition}

\newtheorem*{definition*}{Definition}

\theoremstyle{remark}
\newtheorem{remark}{Remark}
\newtheorem*{remark*}{Remark}

\newcommand{\bDelta}{\mathbf{\Delta}}

\newcommand{\step}{\gamma}

\newcommand{\rate}{R}

\newcommand{\pmax}{P_{\!\max}}
\newcommand{\pc}{P_{\!c}}
\newcommand{\vbound}{V_{0}}
\newcommand{\stochbound}{\hat V_{0}}

\newcommand{\choice}{\boldsymbol{\Pi}}
\newcommand{\smallchoice}{\boldsymbol{\pi}}
\newcommand{\pay}{u}

\newcommand{\breg}{D}

\newcommand{\depth}{\Omega}

\DeclareMathOperator{\ee}{EE}

\newcommand{\pospart}[1]{\left[ #1 \right]_{+}}

\newcommand{\hypref}[1]{\textup{(\hyperref[#1]{H\ref*{#1}})}}

\newcommand{\qstrat}{\boldsymbol{\mathcal{Q}}}
\newcommand{\qsols}{\qstrat^{\ast}}
\newcommand{\strat}{\boldsymbol{\mathcal{X}}}

\newcommand{\qtarget}{\bQ^{\ast}}
\newcommand{\target}{\bX^{\ast}}

\newcommand{\users}{\mathcal{U}}
\newcommand{\user}{u}
\newcommand{\channels}{\mathcal{K}}
\newcommand{\channel}{k}
\newcommand{\chan}{\channel}
\newcommand{\chans}{\channels}
\newcommand{\rx}{N}
\newcommand{\tx}{M}
\newcommand{\effH}{\wilde\bH}

\DeclareMathOperator{\reg}{Reg}
\newcommand{\horizon}{T}

\newcommand{\db}{\textrm{dB}\xspace}
\newcommand{\dbm}{\textrm{dBm}\xspace}
\newcommand{\bps}{\textrm{bps}\xspace}
\newcommand{\Mb}{\textrm{Mb}\xspace}
\newcommand{\joule}{\textrm{J}\xspace}
\newcommand{\meter}{\textrm{m}\xspace}
\newcommand{\km}{\textrm{km}\xspace}
\newcommand{\kmh}{\textrm{km}/\textrm{h}\xspace}
\newcommand{\ms}{\textrm{ms}\xspace}

\newcommand{\hz}{\textrm{Hz}\xspace}
\newcommand{\khz}{\textrm{kHz}\xspace}
\newcommand{\mhz}{\textrm{MHz}\xspace}
\newcommand{\ghz}{\textrm{GHz}\xspace}

\begin{document}


\title{\huge
Learning to Be Green:
Robust Energy Efficiency Maximization in Dynamic MIMO-OFDM Systems}

\author{
Panayotis Mertikopoulos%
,~\IEEEmembership{Member,~IEEE},
and
E. Veronica Belmega%
,~\IEEEmembership{Member,~IEEE}
\thanks{P.~Mertikopoulos is with the French National Center for Scientific Research (CNRS) and the Laboratoire d'Informatique de Grenoble (LIG), F-38000, Grenoble, France.
E. V. Belmega is with ETIS / ENSEA - UCP - CNRS, Cergy-Pontoise, France.}
\thanks{%
This research was supported in part by the European Commission in the framework of the FP7 Network of Excellence in Wireless COMmunications NEWCOM\# (contract no. 318306), by the French National Research Agency projects
NETLEARN (ANR\textendash 13\textendash INFR\textendash 004)
and GAGA (ANR\textendash 13\textendash JS01\textendash 0004\textendash 01) and by ENSEA, Cergy-Pontoise, France.
Part of this was presented at VTC2015-Spring.}
} 

\maketitle

\newacro{BER}{bit error rate}
\newacro{OLE}{online learning}
\newacro{FM}{Foschini\textendash Miljanic}
\newacro{EE}{energy efficiency}
\newacro{AIMD}{additive increase, multiplicative decrease}
\newacro{5G}{fifth generation}
\newacro{SISO}{single-input and single-output}
\newacro{MIMO}{mul\-tiple-input and multiple-output}
\newacro{MUI}{multi-user interference-plus-noise}
\newacro{MAC}{multiple access channel}
\newacro{PMAC}{parallel multiple access channel}
\newacro{CSI}{channel state information}
\newacro{CSIT}{channel state information at the transmitter}
\newacro{BS}{base station}
\newacro{TDD}{time-division duplexing}
\newacro{CDMA}{code division multiple access}
\newacro{FDMA}{frequency division multiple access}
\newacro{DSL}{digital subscriber line}
\newacro{SIC}{successive interference cancellation}
\newacro{SUD}{single user decoding}
\newacro{SINR}{signal-to-interference-and-noise ratio}
\newacro{KKT}{Ka\-rush--Kuhn--Tuc\-ker}
\newacro{WF}{water-filling}
\newacro{IWF}{iterative water-filling}
\newacro{SWF}{simultaneous water-filling}
\newacro{iid}[i.i.d.]{independent and identically distributed}
\newacro{OFDMA}{orthogonal frequency-division multiple access}
\newacro{MXL}{matrix exponential learning}
\newacro{AMXL}[MXL-a]{asynchronous matrix exponential learning}
\newacro{EXL}[MXL-eig]{eigen-based exponential learning}
\newacro{FCC}{Federal Communications Commission}
\newacro{NTIA}{National Telecommunications and Information Administration}
\newacro{GAO}{General Accounting Office}
\newacro{QoE}{quality of experience}
\newacro{QoS}{quality of service}
\newacro{OFDM}{orthogonal frequency division multiplexing}
\newacro{MIMO-OFDM}{multiple-input multiple-output orthogonal frequency division multiplexing}
\newacro{EW}{exponential weight}
\newacro{OGA}{online gradient ascent}
\newacro{OMD}{online mirror descent}
\newacro{APT}{asymptotic pseudotrajectory}
\newacro{ICT}{information and communications technology}
\newacro{MSE}{mean squared error}
\newacro{EPA}{extended pedestrian A}
\newacro{EVA}{extended vehicular A}
\newacro{ETU}{extended typical urban}
\newacro{UL}{uplink}
\newacro{DL}{downlink}
\newacro{CCI}{co-channel interference}

\vspace{-2em}
\begin{abstract}
%
%
In this paper, we examine the maximization of \ac{EE} in next-generation multi-user \acs{MIMO}\textendash\acs{OFDM} networks that evolve dynamically over time \textendash\ e.g. due to user mobility, fluctuations in the wireless medium, modulations in the users' load, etc.
Contrary to the static/stationary regime, the system may evolve in an arbitrary manner so,
targeting a fixed optimum state (either static or in the mean) becomes obsolete;
instead, users must adjust to changes in the system ``on the fly'', without being able to predict the state of the system in advance.
To tackle these issues, we propose a simple and distributed online optimization policy that leads to \emph{no regret}, i.e. it allows users to match (and typically outperform) even the best fixed transmit policy in hindsight, irrespective of how the system varies with time.
Moreover, to account for the scarcity of perfect \ac{CSI} in massive \acs{MIMO} systems, we also study the algorithm's robustness in the presence of measurement errors and observation noise.
Importantly, the proposed policy retains its no-regret properties under very mild assumptions on the error statistics and, on average, it enjoys the same performance guarantees as in the noiseless, deterministic case.
Our analysis is supplemented by extensive numerical simulations which show that, in realistic network environments, users track their individually optimum transmit profile even under rapidly changing channel conditions, achieving gains of up to $600\%$ in \acl{EE} over uniform power allocation policies.

\end{abstract}


\acresetall

\section{Introduction}
\label{sec:intro}

\acused{MIMO}

\IEEEPARstart{T}{he} wildfire spread of Internet-enabled mobile devices and the exponential growth of bandwidth-hungry applications is putting existing wireless systems under enormous strain and is one of the driving forces behind the transition to \ac{5G} mobile networks \cite{ABC+14}.
In this way, the \acs{ICT} industry is faced with a formidable mission:
data rates must be increased significantly in order to meet the soaring demand for wireless broadband, but this task must be accomplished under an extremely tight energy budget.
Thus, to achieve the seamless integration of a diverse set of mobile users, applications and services, current design requirements for \ac{5G} systems target a dramatic decrease in energy-per-bit consumption of the order of $1,000\times$ or more \cite{Qua13,Hua13}.

A contending technology to achieve these design targets is the emerging massive \acs{MIMO} (\acl{MIMO}) paradigm.
Coupled with the use of multiple carrier frequencies via \ac{OFDM}, massive \ac{MIMO} ``goes large'' by employing inexpensive service antennas to focus energy into ever smaller regions of space \cite{HtBD13,RPL+13,LETM14}.
As a result, very large \ac{MIMO} arrays can greatly enhance the reliability of wireless connections and increase throughput and \ac{EE} by a factor of $10\times$ to $100\times$
without requiring the deployment of expensive new air interfaces \cite{LETM14,ABC+14}.
However, due to the massive complexity and variability of such systems, a crucial challenge that arises is that wireless users must also be capable of adapting to a dynamic spectrum landscape ``on the fly'',
usually with minimal coordination and limited information at the device end.

An added challenge in the above considerations is that wireless users often do not have access to perfect \ac{CSI} and \ac{CCI} measurements, especially at the transmitter end \textendash\ for instance, due to pilot contamination in massive \ac{MIMO} systems \cite{LETM14}.
In particular, if the system operates in the presence of uncertainty (imperfect \ac{CSI}, observation noise, etc.), optimization techniques that rely on a greedy, ``one-off'' calculation of optimal transmit characteristics (such as water-filling) are no longer suitable because stochastic fluctuations could lead the system to a suboptimal state \cite{CGM14,MM15}.
On that account, our main objective in this paper will be to provide an adaptive transmit policy for \acl{EE} maximization in dynamic \ac{MIMO}\textendash\ac{OFDM} networks that are subject to uncertainty, measurement errors and/or other unpredictable changes in the wireless medium.


In the general context of \ac{MIMO}\textendash\ac{OFDM} systems, the vast majority of works on \acl{EE} maximization and energy-efficient power allocation have focused on two limit cases \cite{FJLC+13}.
In the static regime \cite{MHL10,MHLT11,ICJF12,GHWC+14,BBMS15,WSV15}, the attributes of the wireless system under study (channel gains, user load, etc.) are assumed effectively static and the system's analysis revolves around techniques from the theory of non-cooperative games and optimization (continuous or discrete).
At the other end of the spectrum, in the ergodic regime \cite{BL11,ICJF12}, the wireless medium is assumed to evolve over a very fast time scale, typically following a sequence of \ac{iid} random variables;
consequently, the figure of merit in problems of this type is the stochastic average of the users' \acl{EE} function.
All these works study the trade-off between the Shannon achievable rate and power consumption either for a single user (via fractional programming) or multiple ones (using the theory of non-cooperative games).
Finally, in the static channel regime, \cite{ZBL05,MPSM05,MCPS06,BLD10,EWD14,HGWLZ12} consider a throughput model that depends on the connection's \ac{BER} and use tools from game theory to characterize the system's stable (equilibrium) states.

In this paper, we focus squarely on dynamic \ac{MIMO}\textendash\ac{OFDM} systems that evolve \emph{arbitrarily} over time (e.g. due to channel variability, fading, mobility, etc.), and we make no statistical hypotheses regarding the dynamics that govern the network's evolution (such as stationarity or ergodicity).
As opposed to the stationary/ergodic regime discussed above, static solution concepts such as Nash/correlated equilibria are no longer relevant because there is no underlying target state to attain (either static or in the mean);
as such, no conclusions can be drawn from the existing literature on energy-efficient power allocation.
Instead, users have to optimize their transmit characteristics on the fly, based only on locally available information of the past state of the system, and hoping to track (or at least emulate the performance of) the \emph{a posteriori} optimum transmit policy.

The most widely used optimization criterion in this setting is that of \emph{regret minimization}, a seminal notion which was first introduced by Hannan \cite{Han57} and which has since given rise to a vigorous literature at the interface of machine learning, optimization, statistics, and game theory \textendash\ for a comprehensive survey, see e.g. \cite{SS11,CBL06,Haz12}.
More precisely, in the language of game theory, a user's (cumulative) regret over a given time horizon is simply the difference between his average payoff (over the time horizon in question) and the payoff that he would have obtained if he had employed the best possible fixed action in hindsight.
Accordingly, in our case, regret minimization corresponds to dynamic transmit policies that are asymptotically optimal in hindsight, irrespective of how the users' effective wireless medium evolves over time.


A regret-based approach was recently employed by the authors of \cite{MagSta14} who studied the problem of power control in infrastructureless wireless networks and proposed an algorithm that minimizes the users' (internal) regret to attain the system's equilibrium.
In a similar vein, \cite{DHK12} studied the transient phase of the \ac{FM} power control algorithm in static environments and used the notion of \emph{swap regret} \cite{BM07} to propose alternative convergent power control schemes;
even more recently, \cite{SMT14} showed that the \ac{FM} dynamics lead to no regret, so they retain their optimality properties in dynamic environments.
Finally, \cite{MB14} and \cite{SMT15} used online optimization techniques and a methodology based on matrix exponential learning \cite{CGM14,MM15,KSST12,TRW05} to derive a no-regret adaptive transmit policy for power control and throughput maximization in cognitive radio networks respectively.
However, the proposed policies drive wireless users to transmit at either full or minimum power (subject to their rate requirements), so they cannot be applied to minimize energy-per-bit consumption in dynamic \ac{MIMO}\textendash\ac{OFDM} systems.

\subsection*{Summary of results and paper outline}

In this paper, we formulate the maximization of \acl{EE} in dynamic \ac{MIMO}--\ac{OFDM} systems as an online semidefinite optimization problem and, drawing on Zinkevich's seminal \acf{OGA} methodology \cite{Zin03}, we propose an adaptive transmit policy which is asymptotically optimal in hindsight \textendash\ i.e. that leads to no regret.
In particular, we show that the proposed algorithm guarantees an $\bigoh(\horizon^{-1/2})$ regret bound after $\horizon$ update epochs (transmission frames), and this bound tightens to $\bigoh(\log\horizon/\horizon)$ if the users' channel gains always remain above a given level.
Furthermore, to address the lack of perfect measurements and \ac{CSIT}, we show that the proposed algorithm retains its optimality properties under very mild statistical hypotheses that are satisfied by the vast majority of error distributions.
Specifically, as long as
\begin{inparaenum}
[\itshape a\upshape)]
\item
there is no \emph{systematic} error in the measurement process;
and
\item
the probability of observing very large errors ($z$) is not higher than $\bigoh(1/z^{2})$,
\end{inparaenum}
the proposed policy leads to no regret and enjoys a mean bound of the same order as in the deterministic case.

The performance of the proposed transmit policy is validated by means of extensive numerical simulations modeling a cellular \ac{OFDMA} network with multiple base stations and mobile \ac{MIMO} users with realistic wireless propagation, fading and mobility features.
Our results show that the proposed policy represents a scalable and flexible method that allows users to attain very high \acl{EE} levels, with gains of up to $600\%$ over uniform/fixed power allocation policies and with surprisingly modest feedback requirements.

Our work here greatly extends our recent conference paper \cite{BM15} where we derived a con\-ti\-nu\-ous-time exponential learning method for \acl{EE} maximization in dynamic \ac{SISO} systems.
Compared to \cite{BM15}, the current paper provides a bona fide learning algorithm for multiple-antenna systems, with discrete-time updates and performance guarantees, and with proven robustness in the presence of uncertainty and observation noise.

The rest of our paper is structured as follows:
in Section \ref{sec:model}, we present our wireless system model and we formulate the problem of dynamic \acl{EE} maximization as an online semidefinite program.
In Section \ref{sec:learning}, we derive our online learning policy, and we establish its no-regret properties and performance guarantees under both perfect and imperfect \ac{CSI}.
Finally, our theoretical analysis is supplemented by extensive numerical simulations in Section \ref{sec:numerics} where we illustrate the gains of the proposed policy under realistic channel gain and mobility conditions.

\section{System Model and Problem Formulation}
\label{sec:model}

Consider a wireless network consisting of several point-to-point connections $\user\in\users = \{1,\dotsc,U\}$ (the system's \emph{users}) that are established over a set of orthogonal subcarriers $\chan\in\chans \equiv \{1,\dotsc,K\}$.
Each connection $\user\in\users$ represents a pair of communicating wireless multi-antenna devices with $\tx_{\user}$ antennas at the transmitter and $\rx_{\user}$ antennas at the receiver.
Thus, focusing on the uplink case, if $\bx_{\chan}^{\user}\in\C^{\tx_{\user}}$ and $\by_{\chan}^{\user}\in\C^{\rx_{\user}}$ denote the signals transmitted and received over connection $\user$ on subcarrier $\chan$, we obtain the familiar baseband signal model:
\begin{equation}
\label{eq:signal-base}
\by_{\chan}^{\user}
	= \bH_{\chan}^{\user\user} \bx_{\chan}^{\user}
	+ \insum_{\user'\neq\user} \bH_{\chan}^{\user'\user} \bx_{\chan}^{\user'}
	+ \bz_{\chan}^{\user},
\end{equation}
where $\bH_{\chan}^{\user'\user} \in \C^{\rx_{\user}\times\tx_{\user'}}$ denotes the transfer matrix between the $\user'$-th transmitter and the $\user$-th receiver over subcarrier $\chan$ while $\bz_{\chan}^{\user}$ is the ambient noise over the channel (including thermal and atmospheric effects, and modeled as a circularly symmetric Gaussian complex vector).
In this way, the \ac{MUI} at the intended receiver of the $\user$-th connection will be:
\begin{equation}
\label{eq:MUI}
\bw_{\chan}^{\user}
	= \insum_{\user'\neq\user} \bH_{\chan}^{\user'\user} \bx_{\chan}^{\user}
	+ \bz_{\chan}^{\user},
\end{equation}
so \eqref{eq:signal-base} may be written more simply as:
\begin{equation}
\label{eq:signal-simplified}
\by_{\chan}^{\user}
	= \bH_{\chan}^{\user\user} \bx_{\chan}^{\user}
	+ \bw_{\chan}^{\user}.
\end{equation}

In the rest of this paper, we will focus on a specific connection $\user\in\users$ and we will treat the \ac{MUI} vector $\bw_{\chan}$ as an aggregate noise variable whose covariance depends on the wireless medium and the transmit characteristics of all other users.
As such, if we drop the user index $\user$ for notational convenience, the signal model \eqref{eq:signal-simplified} attains the more compact form:
\begin{equation}
\label{eq:signal}
\by_{\chan}
	= \bH_{\chan} \bx_{\chan}
	+ \bw_{\chan}.
\end{equation}
Hence, assuming Gaussian input and \ac{SUD} at the receiver, the Shannon rate at the focal connection will be given by the well-known expression \cite{Tel99}:%
\footnote{For the sake of simplicity, constant multiplicative factors such as the bandwidth of the connection have been dropped in \eqref{eq:rate-base}; these factors are reinstated in the numerical analysis of Section \ref{sec:numerics}.}
\begin{equation}
\label{eq:rate-base}
\rate(\bQ)
	= \insum_{\chan\in\chans} \left[
	\log\det\left( \bW_{\chan} + \bH_{\chan} \bQ_{\chan} \bH_{\chan}^{\dag} \right)
	- \log\det \bW_{\chan}
	\right],
\end{equation}
where:
\begin{enumerate}
\item
$\bQ_{\chan} = \ex[\bx_{\chan}\bx_{\chan}^{\dag}] \in \C^{\tx\times\tx}$ is the user's input signal covariance matrix over subcarrier $\chan$.%
\footnote{In the above, expectations are taken over the users' codebooks (assumed Gaussian).}
\item
$\bQ = \diag(\bQ_{1},\dotsc,\bQ_{K})$ is the power profile of the focal user over all subcarriers.
\item
$\bW_{\chan} = \ex[\bw_{\chan} \bw_{\chan}^{\dag}] \in \C^{\rx\times\rx}$ is the \ac{MUI} covariance matrix of the \acl{CCI} plus noise affecting the focal connection (obviously, $\bW_{\chan}$ depends on all other users in the network).
\end{enumerate}

\begin{remark}
The Gaussian input and noise assumptions are fairly standard in the literature:
in particular, Gaussian noise is known to be the worst additive noise distribution with respect to the Shannon achievable rate \cite{diggavi2001worst} while Gaussian input is optimal against a Gaussian environment \cite{Tel99}.
Finally, regarding the decoding technique, \ac{SUD} has the advantage of being simple, distributed, and scalable as it does not require any coordination or signaling among the interfering users.
\end{remark}

In view of the above, if we let
\begin{equation}
\label{eq:channel-eff}
\effH_{\chan}
	= \bW_{\chan}^{-1/2} \bH_{\chan}
\end{equation}
denote the \emph{effective channel matrix} of the focal user over subcarrier $\chan$, the user's Shannon rate \eqref{eq:rate} can be written more concisely as:
\begin{equation}
\label{eq:rate}
\rate(\bQ)
	= \insum_{\chan\in\chans} \log\det\left( \bI + \effH_{\chan} \bQ_{\chan} \effH_{\chan}^{\dag}\right)
	= \log\det\left( \bI + \effH \bQ \effH^{\dag} \right),
\end{equation}
where $\effH = \diag(\effH_{1},\dotsc,\effH_{K})$ is the block-diagonal sum of the user's effective channel matrices over all subcarriers.
Thus, following \cite{MHLT11,ICJF12,GHWC+14,BBMS15}, the user's \acl{EE} function is defined as his Shannon rate per unit of consumed power, i.e.
\begin{equation}
\label{eq:EE}
\ee(\bQ)
	= \frac{\rate(\bQ)}{\pc + \tr(\bQ)}
	= \frac{\log\det\left( \bI + \effH \bQ \effH^{\dag} \right)}{\pc + \tr(\bQ)},
\end{equation}
where $\tr(\bQ) = \sum_{\chan} \tr(\bQ_{\chan})$ is the user's total transmit power while $\pc$ denotes the total power dissipated in all other circuit components of the transmitting device (mixer, frequency synthesizer, digital-to-analog converter, etc.).
This efficiency function (which, formally, has units of bits/Joule) has been widely studied in the literature \cite{CGB04,MCPS06,ICJF12} and it captures the fundamental trade-off between higher spectral efficiency and increased battery life.
Consequently, in the context of power-limited, energy-aware users, we obtain the maximization problem:
\begin{equation}
\label{eq:EE-max}
\begin{aligned}
\text{maximize}
	&\quad
	\ee(\bQ),
	\\
\text{subject to}
	&\quad
	\bQ\in\qstrat,
\end{aligned}
\end{equation}
where
\begin{equation}
\label{eq:qstrat}
\txs
\qstrat
	= \left\{ \diag(\bQ_{1},\dotsc,\bQ_{K}): \bQ_{\chan} \mgeq 0,\;\insum_{\chan} \tr(\bQ_{\chan}) \leq \pmax \right\},
\end{equation}
and $\pmax$ denotes the user's maximum transmit power.

Of course, the user's \acl{EE} function depends not only on the transmitter's signal covariance profile $\bQ$, but also on the transmit characteristics of all other users via the effective channel matrices $\effH_{\chan}$:
in particular, $\effH$ collects all sources of noise and interference that cannot be controlled by the focal transmit/receive pair, so the user's \acl{EE} objective may vary itself over time in an unpredictable way.
On that account, since we wish to focus on dynamic networks that evolve in an arbitrary fashion, we will not be making any specific postulates regarding the behavior of other users in the network and/or the evolution of the user's actual channel matrix $\bH$.
The only generic assumptions that we will make regarding the effective channel matrices $\effH$ are:
\begin{enumerate}
[({A}1)]
\item
$\effH$ remains bounded over the entire transmission horizon (e.g. due to the minimum distance between transmitter and receiver, RF circuit losses, antenna directivity, etc.).
\item
The variability of $\effH$ within each transmission frame is suficiently slow so that the standard caveats of information theory remain valid.
\end{enumerate}

Consequently, if $\effH(t)$ is the user's effective channel matrix at time $t$,
we obtain the following \emph{online} \acl{EE} problem:
\begin{equation}
\label{eq:OEE}
\tag{OEE}
\begin{aligned}
\text{maximize}
	&\quad
	\ee(\bQ;t),
	\\
\text{subject to}
	&\quad
	\bQ\in\qstrat,
\end{aligned}
\end{equation}
where, in obvious notation:
\begin{equation}
\label{eq:EE-time}
\ee(\bQ;t)
	= \frac{\log\det\big( \bI + \effH(t) \bQ \effH^{\dag}(t) \big)}{\pc + \tr(\bQ)}
\end{equation}
denotes the user's energy efficiency function at time $t$.
Thus, given that the user cannot predict the state of the system ahead of time, we will focus on the following sequence of events:
\begin{enumerate}
\item
At each update period $n=1,2,\dotsc$, the user selects a transmit power profile $\bQ(n)\in\qstrat$.
\item
The user's \acl{EE} over the current period is determined by the effective channel matrix $\effH(n)$ at the time of transmission.
\item
At the end of the period, the user selects a new signal covariance profile $\bQ(n+1)$ seeking to maximize his \emph{a priori unknown} objective function $\ee(\bQ;n+1)$, and the process repeats.
\end{enumerate}

Of course, the key challenge in this dynamic framework is that the user does not know ahead of time the effective channel matrix $\effH(n+1)$ that determines his \acl{EE} function at stage $n+1$, so he must try to adapt to the changing network conditions ``on the fly''.
To be sure, if the user had perfect foresight and knowledge of the evolution of $\effH(n)$ in advance, the (fixed) power profile that maximizes the user's average \acl{EE} over a given transmission horizon $\horizon$ would be the solution to the (offline) maximization problem:
\begin{equation}
\label{eq:EE-average}
\max_{\bQ\in\qstrat } \frac{1}{\horizon} \insum_{n=1}^{\horizon} \ee(\bQ;n).
\end{equation}
Obviously however, this ``oracle'' solution cannot be computed without precognitive abilities, so we will focus on adaptive transmit policies $\bQ(n)$ that approach the maximal value of \eqref{eq:EE-average} asymptotically, \emph{irrespective of the system's evolution over time}.

To make this analysis precise, we define the user's (cumulative) \emph{regret} at time $\horizon$ as the cumulative difference between the user's achieved \ac{EE} and the solution of the maximization problem \eqref{eq:EE-average}, i.e. we let:
\begin{equation}
\label{eq:regret}
\reg(\horizon)
	= \max_{\bQ\in\qstrat} \insum_{n=1}^{\horizon} \left[ \ee(\bQ;n) - \ee(\bQ(n);n) \right].
\end{equation}
We then say that a dynamic transmit policy $\bQ(n)$ leads to \emph{no regret} if
\begin{equation}
\label{eq:no-regret}
\limsup_{\horizon\to\infty} \horizon^{-1} \reg(\horizon) \leq 0,
	\quad
	\text{or, equivalently:}
	\quad
	\reg(\horizon) = o(\horizon),
\end{equation}
independently of the evolution of the user's \acl{EE} function.
In this way, a no-regret policy $\bQ(n)$ is \emph{asymptotically optimal in hindsight} in that it provides an asymptotic solution to the average \acl{EE} maximization problem \eqref{eq:EE-average}, without requiring any oracle-like capabilities from the user.

The seminal notion of regret was first introduced in a game-theoretic setting by Hannan \cite{Han57} and it has since given rise to a vast corpus of research at the interface of optimization, statistics and machine learning \textendash\ for a recent survey, see e.g. \cite{CBL06,SS11}.
In particular, if the user's \acl{EE} function does not vary with time (i.e. if the user's effective channels are static), standard arguments from the theory of online optimization \cite{SS11} can be used to show that no-regret policies converge to the set $\qsols = \argmax_{\bQ} \ee(\bQ)$ of maximally energy-efficient power profiles that solve the (static) problem \eqref{eq:EE-max}.
Likewise, if the user could somehow predict an instantaneous optimum policy $\qtarget(n) \in \argmax_{\bQ} \ee(\bQ;n)$ ahead of every stage $n=1,2,\dotsc,\horizon$, we would have $\reg(\horizon) \leq 0$ for all $\horizon$;
by this token, the no-regret requirement \eqref{eq:no-regret} is a crucial indicator that $\bQ(n)$ tracks the optimal solution of \eqref{eq:OEE} as it evolves over time.
The quality of this tracking can be quantified by more sophisticated regret notions such as adaptive \cite{HS09} or shifting \cite{CBGLS12} regret.
We focus here on the simpler case of external regret minimization due to space limitations;
however, in Section \ref{sec:numerics}, we explore this issue via extensive numerical simulations.

\begin{remark*}
We should also note here that the no-regret property \eqref{eq:no-regret} is a ``worst-case'' guarantee that carries no assumptions on the evolution of the user's environment over time:
the user's channels could evolve randomly (following some stationary, ergodic process, as in the case of fast-fading), adversarially (e.g. if the user is subject to jamming), or not at all (in the static regime).
As such, in the special case where the wireless medium is affected only by the behavior of other users in the network, a natural question that arises is whether the use of a no-regret policy by all users leads to an equilibrium of the underlying game.%
\footnote{For instance, it is well-known that internal regret minimization implies convergence to the set of correlated equilibria \cite{HMC00}.}
We address this issue in more detail in Section \ref{sec:numerics}.
\end{remark*}

\section{Online Learning}
\label{sec:learning}

A first idea to achieve no regret in the online \acl{EE} maximization problem \eqref{eq:OEE} would be to calculate at each stage the power profile that maximizes \acl{EE} based on the latest available information at the previous stage.
However, as can be seen by a standard online optimization argument, this policy may lead to \emph{positive} regret:
for instance, when the user's channel alternates every other period between two values \textendash\ say $\bH_{a}$ and $\bH_{b}$ with corresponding optimal power profiles $\qtarget_{a}$ and $\qtarget_{b}$ \textendash\ best-responding to the last observed system state performs strictly worse than the \emph{fixed} policy $(\qtarget_{a} + \qtarget_{b})/2$ \cite{SS11}.
With this in mind, we propose in this section an \emph{adaptive} power allocation policy that utilizes \emph{all} past information in a recursive way based on Zinkevich's seminal \ac{OGA} method \cite{Zin03}.

For simplicity, we first consider the case where the transmitter has access to perfect \ac{CSI} and \ac{MUI} measurements and we derive an anytime $\bigoh(\horizon^{1/2})$ bound for the user's regret;
we then show that this bound can be tightened to $\bigoh(\log\horizon)$ if the user's effective channels always remain above a certain threshold and the algorithm's step-size is chosen accordingly.
The robustness of these guarantees in the presence of noise and uncertainty is then discussed in Section \ref{sec:imperfectCSI}.

\subsection{Energy efficiency maximization as an online concave problem}
\label{sec:reformulation}

The first difficulty in designing a no-regret policy for the online fractional program \eqref{eq:OEE} is that the user's \acl{EE} function is not concave.
This is perhaps most easily seen in the \ac{SISO} case where the user's \acl{EE} objective becomes:
\begin{equation}
\label{eq:EE-simple}
\ee(\bp)
	= \frac{\sum_{\chan}\log(1+\tilde g_{\chan} p_{\chan})}{\pc + \sum_{\chan} p_{\chan}},
\end{equation}
where $\bp = (p_{1},\dotsc,p_{K})$ denotes the user's power allocation vector and $\tilde g_{\chan}$ is the effective channel gain of channel $\chan$.
Clearly, the fractional objective \eqref{eq:EE-simple} is not concave with respect to any $p_{\chan}$;
however, $\ee(\bp)$ can be recast as a concave function by employing the so-called Charnes\textendash Cooper transformation \cite{CC62} for turning fractional programs into concave ones.%
\footnote{See also \cite{ICJF12} for a similar use of the Charnes\textendash Cooper transformation in the context of \acl{EE} maximization.}
Specifically, if we set
\begin{equation}
\label{eq:CC-simple}
\txs
x_{0}
	= (\pc + \sum_{\chan} p_{\chan})^{-1},
	\quad
\bx
	= x_{0}\cdot\bp,
\end{equation}
we readily obtain $\ee(\bp) = x_{0} \sum_{\chan} \log(1 + \tilde g_{\chan} x_{\chan}/x_{0})$,
and this last function is concave
because the summands $x_{0} \log(1 + \tilde g_{\chan} x_{\chan}/x_{0})$ are jointly concave in $x_{0}$ and $x_{\chan}$.
We may then get rid of the parameter $x_{0}$ by noticing that $x_{0} = \frac{1}{\pc}(1 - \sum_{\chan} x_{\chan})$ which restricts the energy efficiency in $(x_{0}, x_{\chan})$ to an affine set on which it remains concave.  Thus, by rewriting $\bx$ as $\bx = \bp/(\pc + \sum_{\chan} p_{\chan})$, solving for $\bp$ and substituting in $\ee(\bp)$ to obtain a concave reformulation of \eqref{eq:EE-simple}.

In the general \ac{MIMO} framework, this procedure amounts to the change of variables:
\begin{equation}
\label{eq:Xdef}
\bX
	= \frac{\pc + \pmax}{\pmax} \frac{\bQ}{\pc + \tr(\bQ)},
\end{equation}
where we have introduced the normalization constant $(\pc + \pmax)/\pmax$ in order to have $\tr(\bX) \leq 1$ for all $\bQ\in\qstrat$ (with equality if and only if $\tr(\bQ) = \pmax$).
Solving for $\bQ$ then yields
\begin{equation}
\label{eq:QfromX}
\bQ
	= \frac{\pc \pmax}{\pc + \pmax(1 - \tr(\bX))} \bX,
\end{equation}
so, after substituting in \eqref{eq:EE}, we obtain the maximization objective
\begin{equation}
\label{eq:pay}
\pay(\bX)
	= \ee(\bQ)
	= \frac{\pc + \pmax (1 - \tr(\bX))}{\pc(\pc + \pmax)}
	\log\det\left(\bI + \frac{\pc \pmax \effH \bX \effH^{\dag}}{\pc + \pmax(1 - \tr(\bX))} \right),
\end{equation}
while the corresponding feasible region of \eqref{eq:EE-max} attains the simple form:
\begin{equation}
\label{eq:strat}
\txs
\strat
	= \left\{\diag(\bX_{1},\dotsc,\bX_{\chan}): \bX_{\chan} \mgeq0 \text{ and } \insum_{\chan} \tr(\bX_{\chan}) \leq 1 \right\}.
\end{equation}

Given that $\rate(\bQ)$ is concave in $\bQ$, the function $F(\bX,x) = \frac{\pmax}{\pc+\pmax} x\cdot\rate(\bX/x)$ will be jointly concave in $\bX$ and $x$ \cite{BV04}, so $\pay(\bX)$ will also be concave in $\bX$ as the restriction of $F(\bX,x)$ to the convex set $\pc \pmax x = \pc + \pmax (1 - \tr(\bX))$.
In this way, \eqref{eq:OEE} boils down to the online concave maximization problem:
\begin{equation}
\label{eq:OP}
\begin{aligned}
\text{maximize}
	&\quad
	\pay(\bX;n),
	\\
\text{subject to}
	&\quad
	\bX\in\strat,
\end{aligned}
\end{equation}
where, as before, the dependence on $n = 1,2\dotsc$, reflects the evolution of the user's effective channel matrices over time.
Thus, in view of all this, we will first derive a no-regret transmit policy $\bX(n)$ for the online concave problem \eqref{eq:OP} and we will then use the inverse transformation \eqref{eq:QfromX} to obtain a no-regret policy for \eqref{eq:OEE}.

\subsection{Learning with perfect \ac{CSI}}
\label{sec:perfectCSI}

Building on Zinkevich's \acl{OGA} method \cite{Zin03}, the core idea of our approach will be to track the gradient matrix $\bV = \nabla \pay$ of the user's (time-varying) utility function and then project back to the problem's feasible region when the user's power constraints are violated.
To that end, some straightforward matrix calculus yields:
\begin{equation}
\label{eq:V}
\bV
	= \nabla \pay
	= \frac{\pmax}{\pc + \pmax} \left[
	\bA + \frac{\tr(\bA \bQ) - \rate(\bQ)}{\pc} \cdot \bI
	\right],
\end{equation}
where
$\bQ$ is calculated in terms of $\bX$ via \eqref{eq:QfromX}
and
\begin{equation}
\label{eq:dR}
\bA
	\equiv \nabla\rate(\bQ)
	= \effH^{\dag} \big[ \bI + \effH \bQ \effH^{\dag} \big]^{-1} \effH.
\end{equation}
The above expression shows that $\bV$ can be calculated at the transmitter as a function of the connection's effective channel matrix $\effH$ (which, in turn, can be estimated at the receiver end and then fed back to the transmitter via a dedicated backbone channel or as part of a \acs{TDD} downlink subframe).
Moreover, since $\bV$ is a bounded function of $\effH$ and the channel matrices $\effH(n)$ are assumed bounded, the induced sequence of gradient matrices $\bV(n) \equiv \nabla\pay(\bX(n);n)$ will also be bounded.
We will therefore assume that there exists a constant $\vbound>0$ such that
\begin{equation}
\label{eq:V-bound}
\norm{\bV(n)}
	\leq \vbound
	\quad
	\text{for all $n=1,2,\dotsc,$}
\end{equation}
where $\norm{\bV} = \tr(\bV^{\dag}\bV)^{1/2}$ denotes the Frobenius (matrix) norm of $\bV$.

In view of the above, and assuming for the moment perfect knowledge of $\bV(n)$ at the transmitter, we will consider the matrix-valued \acl{OGA} scheme:
\begin{equation}
\label{eq:OGA}
\tag{OGA}
\begin{aligned}
\bX(n+1)
	&= \choice(\bX(n) + \step_{n} \bV(n)),
\end{aligned}
\end{equation}
where $\step_{n}>0$ is a nonincreasing step-size sequence and $\choice$ denotes the matrix projection map:
\begin{equation}
\label{eq:choice}
\txs
\choice(\bY)
	= \argmin\nolimits_{\bX\in\strat} \norm{\bX - \bY}^{2}.
\end{equation}
As we show in Appendix \ref{app:proofs-projection}, the matrix projection $\choice(\bY)$ can be calculated by the simple expression:
\begin{equation}
\label{eq:choice-explicit}
\choice(\bY)
	= \bU \cdot \diag(\smallchoice(\by)) \cdot \bU^{\dag},
\end{equation}
where the tuple $(\by,\bU)$ diagonalizes $\bY$ (i.e. $\bY = \bU\cdot\diag(\by) \cdot \bU^{\dag}$) and
\begin{equation}
\label{eq:smallchoice}
\pi_{i}(\by)
	= \begin{cases}
	0
		&\text{if $y_{i} < 0$},
		\\
	y_{i}
		&\text{if $y_{i} \geq 0$ and $\sum_{j} [y_{j}]_{+} < 1$},
		\\
	\pospart{y_{i} - \lambda}
		&\text{if $y_{i} \geq 0$ and $\sum_{j}[y_{j}]_{+} \geq 1$},
	\end{cases}
\end{equation}
with $\lambda>0$ chosen so that $\sum_{i:y_{i}\geq0} \pospart{y_{i} - \lambda} = 1$.%
\footnote{Recall here that $\bY(n)$ is Hermitian (because $\bV(n)$ is Hermitian for all $n$), so its eigenvalues are real.
Just as in water-filling methods \cite{YRBC04}, the Lagrange multiplier $\lambda>0$ can then be calculated by sorting $\by$ and performing a line search for $\lambda$.}

\begin{figure}[t]
  \centering
\begin{tikzpicture}
[scale=1.5]

\draw [thick,pattern = north east lines, pattern color = black!10] (45:3) arc (45:-45:3);
\draw node at (2.4,-1.4) {\large$\strat$};

\node (xn) at (25:2.5) {\small\textbullet};
\node [above right] at (25:2.5) {$\bX_{n}$};

\node (an1) at (5:5) {\small\textbullet};
\node (xn1) at (5:3) {\small\textbullet};
\node [above left] at (5:3) {$\bX_{n+1}$};

\node (xn22) at (-15:3) {\small\textbullet};
\node[above left] at (-15:3) {$\bX_{n+2}$};
\node (an22) at (-15:4) {\small\textbullet};

\draw[-stealth] (xn) -- (an1) node [near end,above]{$+\step_{n}\bV_{n}$};
\draw [-stealth] (xn1) -- (an22) node[near end, right]{\;$+\step_{n+1}\bV_{n+1}$};

\draw [-stealth, densely dashed] (an1) -- (xn1) node [midway,below]{$\choice$};
\draw[-stealth, densely dashed] (an22) -- (xn22) node [midway, below] {$\choice$};


\end{tikzpicture}

\caption{Schematic representation of the recursive learning scheme \eqref{eq:OGA}.}
\vspace{-3ex}
\label{fig:OGA}
\end{figure}
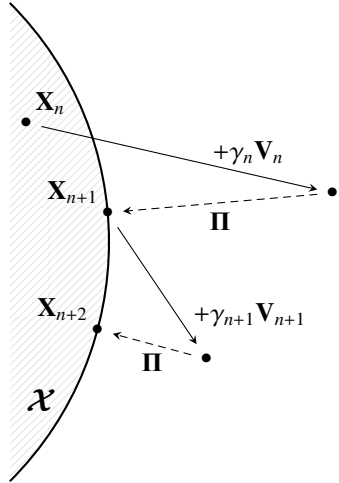

The iterative process \eqref{eq:OGA} will be the main focus of our paper, so we proceed with some remarks  (see also Fig.~\ref{fig:OGA} for a schematic representation and Alg.~\ref{alg:OGA} for a pseudocode version):

\paragraph{Implementation properties}
From a practical point of view, \eqref{eq:OGA} has the following desirable properties:
\begin{enumerate}
[(P1)]
\item
\emph{Distributedness:}
users require the same information as in distributed water-filling \cite{YRBC04,SPB09-sp}.
\item
\emph{Statelessness:}
users do not need to know the state of the system (e.g. the number of users in the network or its topology).
\item
\emph{Reinforcement:}
users tend to become more energy-efficient based on their past observations.
\item
\emph{Asynchronicity:}
the algorithm does not require a global update timer or any further signaling/coordination between users.
\end{enumerate}

\paragraph{Computational complexity}
From a computational standpoint (which is crucial in massive \ac{MIMO} systems), each iteration of Algorithm \ref{alg:OGA} requires a number of elementary binary operations which is polynomial (with a low degree) on the number of transmit/receive antennas and the number of subcarriers.
Specifically, letting $S = \max\{\tx,\rx\}$ and recalling that $\effH$ and $\bQ$ consist of $K$ diagonal blocks, the required matrix multiplication and inversion steps for $\bA$ and $\bV$ carry a complexity of $\bigoh(K S^{\omega})$, with the complexity exponent $\omega$ being as low as $2.373$ if fast Coppersmith\textendash Winograd multiplication methods are employed \cite{DS13}.
As for the projection step $\bX = \choice(\bY)$, Eqs. \eqref{eq:choice-explicit} and \eqref{eq:smallchoice} show that it can also be carried out in $\bigoh(KS^{\omega})$ operations:
the diagonalization in \eqref{eq:choice-explicit} involves $\bigoh(KS^{\omega})$ steps while \eqref{eq:smallchoice} only requires $\bigoh(KM)$ operations for calculating the projection to the simplex \cite{MdP89}.

\begin{algorithm}[t]
{%
\sf
\vspace{2pt}
Parameter:
variable step-size sequence $\step_{n}>0$.
\\
Initialize:
$n \leftarrow 0$;\;
$\bX \leftarrow 0$.
\\
\Repeat
{%
$n \leftarrow n+1$;
\\
\Comment{Pre-transmission phase: set signal covariance matrix}
\\
$\bQ
	\leftarrow \pc \pmax / (\pc + \pmax(1 - \tr\bX)) \cdot \bX$;
\\
\textbf{transmit;}
\\
\Comment{Post-transmission phase: receive feedback and update}
\\
get $\effH$;
\\
$\bA \leftarrow \effH^{\dag} \big[ \bI + \effH \bQ \effH^{\dag} \big]^{-1} \effH$;
\\
$\bV \leftarrow
	\pmax / (\pc + \pmax) \, \left( \bA + [\tr(\bA \bQ) - \rate(\bQ))/\pc \cdot \bI \right]$;
\\
$\bX \leftarrow \choice(\bX + \step_{n}\bV)$;
\\
\textbf{until} transmission ends.
} 
} 
\caption{Online gradient ascent (\acs{OGA}) for dynamic energy efficiency maximization.}
\label{alg:OGA}
\end{algorithm}

\smallskip

With all this in mind, our main result for \eqref{eq:OGA} is as follows:

\begin{theorem}
\label{thm:reg-bound}
Assume that \eqref{eq:OGA} is run with a variable step-size $\step_{n}$ such that $\step_{n} \to 0$ and $n\step_{n} \to \infty$.
Then, the induced transmit policy $\bQ(n)$ leads to no regret in the online \acl{EE} maximization problem \eqref{eq:OEE};
specifically, \eqref{eq:OGA} enjoys the cumulative regret bound:
\begin{equation}
\label{eq:reg-bound}
\reg(\horizon)
	\leq \frac{1}{\step_{\horizon}} + \frac{1}{2} \vbound^{2} \sum_{n=1}^{\horizon} \step_{n},
\end{equation}
or, using a step-size sequence of the form $\step_{n} = \step n^{-1/2}$:
\begin{equation}
\label{eq:reg-sqrt}
\reg(\horizon)
	\leq \frac{1 + \step^{2}\vbound^{2}}{\step} \sqrt{\horizon}.
\end{equation}
\end{theorem}

\begin{IEEEproof}
See Appendix \ref{app:proofs-reg-bound}.
\end{IEEEproof}

The anytime regret bound \eqref{eq:reg-bound} will be our core performance guarantee for Algorithm \ref{alg:OGA}, so some remarks are in order:

\setcounter{paragraph}{0}

\paragraph{Fine-tuning $\step_{n}$}
Theorem \ref{thm:reg-bound} shows that taking $\step_{n} \propto n^{-\alpha}$ for some $\alpha\in(0,1)$ leads to a regret guarantee that is $\bigoh(\horizon^{\omega})$ with $\omega = \max\{\alpha,1-\alpha\}$;%
\footnote{To see this, simply note that $\sum_{n=1}^{\horizon} n^{-\alpha} = \bigoh(\horizon^{1-\alpha})$ for large $\horizon$ and $\alpha\in(0,1)$.}
as such, \eqref{eq:reg-sqrt} captures the optimal asymptotic behavior of the bound \eqref{eq:reg-bound} for step-size sequences of the form $\step_{n} = \step/n^{\alpha}$.
In fact, if $\vbound$ can be estimated by the transmitter beforehand, the step-size parameter $\step$ can be fine-tuned further in order to minimize the coefficient of $\horizon^{1/2}$ in \eqref{eq:reg-sqrt}.
Doing just that gives $\step = 1/\vbound$ and provides the optimized bound:
\begin{equation}
\label{eq:reg-bound-opt}
\reg(\horizon)
	\leq 2 \vbound \sqrt{\horizon}.
\end{equation}
Since $\vbound$ is a bound on the Frobenius norm of the block-diagonal gradient matrices $\bV(n)$, the guarantee \eqref{eq:reg-bound-opt} becomes $\bigoh(K\tx^{2})$ so the algorithm's overall regret will be at most linear in the number of subcarriers and quadratic in the number of antennas.
This guarantee is key for massive \ac{MIMO} systems (where the number of transmit/receive antennas can grow to be quite large) because it provides a worst-case estimate for the system's equilibration time.
That being said, \eqref{eq:reg-bound-opt} only becomes tight in adversarial environments (e.g. in the presence of jamming);
in typical scenarios, the user's regret usually decays much faster and the system attains a stable, no-regret state within a few iterations, even for large numbers of antennas per user \textendash\ cf. the detailed discussion in Sec.~\ref{sec:numerics}.

\paragraph{The static case}
If the user's effective channels remain static over time and $\bQ(n)$ is a no-regret policy, a straightforward concavity argument can be used to show that $\max_{n} \ee(\bQ(n))$ converges to the solution of the (static) \ac{EE} maximization problem \eqref{eq:EE-max} \cite{SS11,KM14}.
In this way, \eqref{eq:OGA} can also be seen as a provably convergent low-cost algorithm for solving \eqref{eq:EE-max};
furthermore, as we show in what follows, this convergence result continues to hold even in the presence of imperfect \ac{CSIT} and measurement errors.

\paragraph{Initialization}
The agnostic initialization $\bX(0) = 0$ of Algorithm \ref{alg:OGA} means that the focal transmitter remains effectively silent during the first transmission frame (recall that $\bQ \propto \bX$).
As such, the first iteration of \eqref{eq:OGA} can be seen as a ``handshake'' that allows the transmitter to estimate his effective wireless medium before starting the bona fide transmission of data frames.
If the transmitter begins with a given belief regarding his effective channel conditions, the algorithm can be initialized more aggressively in a manner consistent with the user's initial expectations (setting for instance $\bX = (K\tx)^{-1} \bI$ for uniform power allocation across subcarriers and antennas).
In so doing, the regret bound \eqref{eq:reg-bound} can be tightened further but this only makes a significant difference if the transmission horizon $\horizon$ is very short (in the order of a few frames).

\paragraph{Logarithmic regret under fair channel conditions}
\label{sec:log-regret}
As stated, Theorem \ref{thm:reg-bound} provides a worst-case guarantee which holds without any further caveats on the evolution of the channels from one stage to the next (other than basic information-theoretic hypotheses that allow the receiver to decode the transmitter's signal).
As such, another important question that arises is whether we can achieve stronger performance guarantees under the additional hypothesis that channel conditions do not become too bad.

To quantify this, note first that the Shannon rate function $\rate(\bQ) = \log\det(\bI + \effH \bQ \effH^{\dag})$ is strongly concave in $\bQ$ with a strong concavity constant that is an increasing function of the singular values of $\effH$.%
\footnote{Recall here that a function $f$ is strongly concave with constant $c>0$ if $\hess(f) \mleq -c\bI$.}
Accordingly, since the user's energy efficiency function $\ee(\bQ)$ can be expressed as a perspective transformation $\rate(\bQ)\mapsto x \rate(\bX/x)$, the same will also hold for the strong concavity constant of $\pay(\bX)$ over $\strat$ \cite{BV04,KSST12}.
On that account, if we assume that:
\begin{equation}
\label{eq:pay-strong}
\hess(\pay(\bX;n))
	\leq - a\,\bI
	\quad
	\text{for some $a>0$ and for all $n=1,2,\dotsc$, $\bX\in\strat$,}
\end{equation}
we obtain the following stronger result:

\begin{proposition}
\label{prop:reg-log}
Assume that \eqref{eq:OGA} is run with the step-size sequence $\step_{n} = \step/n$ for some $\step\geq a^{-1}$ with $a$ as in \eqref{eq:pay-strong}.
Then, the induced transmit policy $\bQ(n)$ enjoys the logarithmic regret bound:
\begin{equation}
\label{eq:reg-log}
\reg(\horizon)
	\leq \frac{1}{2} \step \vbound^{2} (1 + \log\horizon).
\end{equation}
\end{proposition}

\begin{IEEEproof}
See Appendix \ref{app:proofs-reg-bound}.
\end{IEEEproof}

Proposition \ref{prop:reg-log} provides us with an important rule of thumb for choosing the step-size sequence of Algorithm~\ref{alg:OGA}.
On the one hand, if the user expects that his effective channel can become arbitrarily bad (e.g. due to network congestion or deep fading events), the \ac{OGA} algorithm should be run with a $n^{-1/2}$ step-size sequence that allows higher adaptability to strongly varying channel conditions.
Otherwise, if the user expects reasonable channel quality over his transmission horizon, the ``softer'' step-size choice $\step_{n}\propto n^{-1}$ minimizes the danger of overcompensating for transmit directions that appear suboptimal and allows the user to converge to a no-regret state faster.

\subsection{Learning under uncertainty}
\label{sec:imperfectCSI}

A key assumption in our analysis so far is that the transmitter has access to perfect \ac{CSI} and \ac{MUI} measurements with which to calculate the gradient matrices $\bV(n)$ at each stage;
in practice however, factors such as pilot contamination, sparse feedback and imperfect channel sampling could have a deleterious effect on the algorithm's performance.
As such, our goal in this section will be to analyze \eqref{eq:OGA} in the presence of uncertainty and measurement errors.

To formalize this, we will assume that, at each stage $n=1,2,\dotsc$, the transmitter observes a noisy estimate $\hat\bV(n)$ of $\bV(n)$ satisfying the statistical hypotheses:
\begin{enumerate}
[({H}1)]
\item
\label{hyp:zeromean}
\emph{Unbiasedness:}
\begin{equation}
\tag{H1}
\label{eq:zeromean}
\ex\big[ \hat\bV(n) \given \bQ(n-1) \big]
	= 0.
\end{equation}

\item
\label{hyp:tailbound}
\emph{Tame error tails:}
\begin{equation}
\tag{H2}
\label{eq:tailbound}
\prob\left( \smallnorm{\hat\bV(n) - \bV(n)} \geq z \right)
	\leq B/z^{\beta}
	\quad
	\text{for some $B>0$ and for some $\beta>2$.}
\end{equation}
\end{enumerate}
Clearly, both hypotheses are quite mild from a practical point of view.
First, the unbiasedness hypothesis \eqref{eq:zeromean} simply amounts to asking that there is no \emph{systematic} error in the user's measuremernts.
Likewise, Hypothesis \eqref{eq:tailbound} is a bare-bones assumption on the probability of observing very high errors and is satisfied by the vast majority of statistical error distributions (including uniform, Gaussian, log-normal, and all Lévy-type error processes);
in particular, we will \emph{not} be assuming that the error process $\bZ(n) = \hat\bV(n) - \bV(n)$ is \ac{iid}, state-independent, or even a.s. bounded.

Remarkably, under these minimal hypotheses, we have:

\begin{theorem}
\label{thm:reg-stoch}
Assume that \eqref{eq:OGA} is run with noisy measurements $\hat\bV(n)$ satisfying Hypotheses \eqref{eq:zeromean} and \eqref{eq:tailbound}, and with a variable step-size sequence of the form $\step_{n} = \step/n^{\alpha}$ for some $\alpha\in (2/\beta,1)$.
Then, the induced transmit policy $\bQ(n)$ leads to no regret \textup(a.s.\textup) and enjoys the mean regret bound:
\begin{equation}
\label{eq:reg-mean}
\ex[\reg(\horizon)]
	\leq \frac{1}{\step_{\horizon}} + \frac{\stochbound^{2}}{2} \sum_{n=1}^{\horizon} \step_{n},
\end{equation}
where $\stochbound^{2} = \sup_{n} \ex\big[ \smallnorm{\hat\bV(n)}^{2} \big]$.
\end{theorem}


Theorem \ref{thm:reg-stoch} (proven in Appendix \ref{app:proofs-reg-stoch}) will be our main result in the context of dynamic \acl{EE} maximization under imperfect \ac{CSI}, so a few remarks are in order:

\setcounter{paragraph}{0}

\paragraph{Step-sizes vs. large error probabilities}
The requirement $\alpha\beta>2$ of Theorem \ref{thm:reg-stoch} indicates a trade-off between the probability of observing very large errors and achieving low regret \eqref{eq:reg-mean}.
Specifically, if the error distribution of $\bZ(n) = \hat\bV(n) - \bV(n)$ has very heavy tails (i.e. \eqref{eq:tailbound} does not hold for $\beta\gg2$), Algorithm \ref{alg:OGA} must be bootstrapped with a conservative step-size sequence $\step_{n} \propto 1/n^{\alpha}$ for some $\alpha\approx 1$;
in so doing however, the first term of \eqref{eq:reg-mean} becomes almost linear, so the user might experience relatively high regret on average (due to the high probability of observing very large errors).
On the other hand, if the tails of $\hat\bV(n)$ are lighter (for instance, the standard case of normally distributed errors exhibits exponentially thin tails, so \eqref{eq:tailbound} holds for all $\beta$), Algorithm \ref{alg:OGA} can be employed with a more adaptive step-size sequence that guarantees a lower regret bound.

In particular, if \eqref{eq:tailbound} holds for some $\beta>4$, \eqref{eq:OGA} can be used with a step-size sequence of the form $\step_{n} = \step n^{-1/2}$ which achieves the optimal behavior of \eqref{eq:reg-mean}, viz.
\begin{equation}
\label{eq:reg-mean-sqrt}
\ex[\reg(\horizon)]
	\leq \frac{1 + \step^{2} \stochbound^{2}}{\step} \sqrt{\horizon}.
\end{equation}
Thus, if the mean square bound $\stochbound^{2} = \sup_{n} \ex\big[ \smallnorm{\hat\bV(n)}^{2} \big]$ can be estimated ahead of time,%
\footnote{Note here that $\stochbound$ is guaranteed to be finite on account of Hypothesis \eqref{eq:tailbound}.}
the step-size sequence $\step_{n}$ can be optimized further.
More precisely, working as in the deterministic case, the coefficient of $\horizon^{1/2}$ in \eqref{eq:reg-mean-sqrt} is minimized when $\step = 1/\stochbound$, so we obtain the optimized bound:
\begin{equation}
\label{eq:reg-mean-opt}
\ex[\reg(\horizon)]
	\leq 2 \stochbound \sqrt{\horizon}.
\end{equation}

\paragraph{The estimation process}
The no-regret properties of \eqref{eq:OGA} under uncertainty rely on the availability of statistically unbiased measurements $\hat\bV$ of $\bV$.
In turn, given that users have perfect knowledge of their individual transmit covariance matrices, this requirement boils down to constructing an unbiased estimator of the matrix $\bA = \effH^{\dag} \big( \bI + \effH \bQ \effH^{\dag} \big)^{-1} \effH$.
In our Gaussian context, this can be accomplished via the statistical sampling process of \cite{CGM14,MM15} which provides an unbiased estimator of $\bA$ with exponentially decaying error tails (i.e. \eqref{eq:tailbound} holds for all $\beta>2$).
However, due to space limitations we will not address this question in more detail here.

\paragraph{Fair channel conditions and noise}
As before, the regret guarantee \eqref{eq:reg-mean} can be tightened significantly if the user's effective channel conditions satisfy \eqref{eq:pay-strong}.
In that case, running \eqref{eq:OGA} with step-sizes $\step_{n}\propto 1/n$, we obtain the following stochastic analogue of Proposition \ref{prop:reg-log}:

\begin{proposition}
\label{prop:reg-log-stoch}
With notation as in Theorem \ref{thm:reg-stoch}, assume that \eqref{eq:OGA} is run with noisy measurements and a variable step-size sequence $\step_{n} = \step/n$ for some $\step\geq a^{-1}$ with $a$ defined as in \eqref{eq:pay-strong}.
Then, the induced transmit policy $\bQ(n)$ leads to no regret \textup(a.s.\textup) and enjoys the mean guarantee:
\begin{equation}
\label{eq:reg-log-mean}
\ex[\reg(\horizon)]
	\leq \frac{1}{2} \step \stochbound^{2} (1 + \log\horizon).
\end{equation}
\end{proposition}

\begin{IEEEproof}
See Appendix \ref{app:proofs-reg-stoch}.
\end{IEEEproof}

As in the perfect \ac{CSI} case, Proposition \ref{prop:reg-log-stoch} provides a rule of thumb for achieving lower regret faster when the user's (effective) wireless medium is not too bad:
as long as \eqref{eq:pay-strong} holds for some $a>0$, the user can achieve logarithmic regret, even with very noisy measurements.

\section{Numerical Results}
\label{sec:numerics}

\begin{table}[t]
\caption{wireless network simulation parameters}
\label{tab:parameters}
\vspace{-1em}
\centering
\footnotesize
\begin{tabular}{|c|c||c|c|}
\hline
\textbf{Parameter}
	&\textbf{Value}
	&\textbf{Parameter}
	&\textbf{Value}
	\\
	\hline
Number of cells
	&$19$ (hexagonal)
	&Cell radius
	&$1\,\km$
\\
\hline
User density
	&$500\,\textrm{users}/\km^{2}$
	&Time frame duration
	&$5\,\ms$
\\
\hline
Propagation model
	&COST Hata
	&BS/MS antenna height
	&$32\,\meter$ / $1.5\,\meter$
\\
\hline
Central frequency
	&$2.5\,\ghz$
	&Total bandwidth
	&$11.2\,\mhz$
\\
\hline
\acs{OFDM} subcarriers
	&$1024$
	&Subcarrier spacing
	&$11\,\khz$
\\
\hline
Spectral noise density ($20\,^{\circ}\textrm{C}$)
	&$-174\,\dbm/\hz$
	&User speed
	&$[3,130]\,\kmh$
\\
\hline
Maximum transmit power
	&$\pmax = 33\,\dbm$
	&Non-radiative power
	&$\pc = 20\,\dbm$
\\
\hline
Transmit antennas per device
	&$\tx=4$
	&Receive antennas per link
	&$\rx=8$
\\
\hline
\end{tabular}
\vspace{-2em}
\end{table}

In this section, we assess the practical performance aspects of the \ac{OGA} algorithm via numerical simulations.
For presentational clarity, we only present here a representative subset of these results but our conclusions apply to a wide range of wireless network parameters and specifications.

Our setup is as follows:
we consider a cellular \ac{OFDMA} wireless network occupying a $10\,\mhz$ band divided into $1024$ subcarriers around a central frequency of $f_{c} = 2.5\,\ghz$.
Wireless signal propagation is modeled following the well-known COST Hata model \cite{Hat80,COST99} and the spectral noise density is taken to be $-174\,\dbm/\hz$ at $20\,^{\circ}\textrm{C}$ (for a detailed overview of simulation parameters, see Table \ref{tab:parameters}).
Network coverage is provided by $19$ hexagonal cells (each with a radius of $1\,\km$) that form a honeycomb pattern spanning an urban area with wireless user density $\rho = 500\,\textrm{users}/\km^{2}$.
To minimize complexity, \ac{OFDM} subcarriers are allocated to wireless users in each cell following a simple randomized access scheme that assigns different users to disjoint subcarrier sets \cite{SZTK08};
as such, the main sources of \ac{CCI} are connections in neighboring cells that utilize the same subcarriers.

To model this, we focus on a set of $U=15$ transmitting users that are located in different cells (following a Poisson point process sampling) and share $K=8$ common subcarriers.
Each wireless transmitter is further assumed to have $\tx=4$ transmit antennas, a maximum transmit power of $\pmax = 40\,\dbm$ and circuit (non-radiative) power consumption of $\pc = 20\,\dbm$;
at the receiver end, we consider $\rx=8$ receive antennas per connection and a receiver noise figure of $7\,\db$.
Finally, communication occurs over a \ac{TDD} scheme with frame duration $T_{f} = 5\,\ms$:
transmission takes place during the \ac{UL} subframe while the receivers process the received signal and provide feedback during the \ac{DL} subframe;
upon reception of the feedback, the users update their power profiles following Alg.~1 and the process repeats.

\begin{figure*}[t]
\footnotesize
\subfigure[Transmit power evolution under \ac{OGA}]{\label{fig:static-power}%
\includegraphics[width=.48\textwidth]{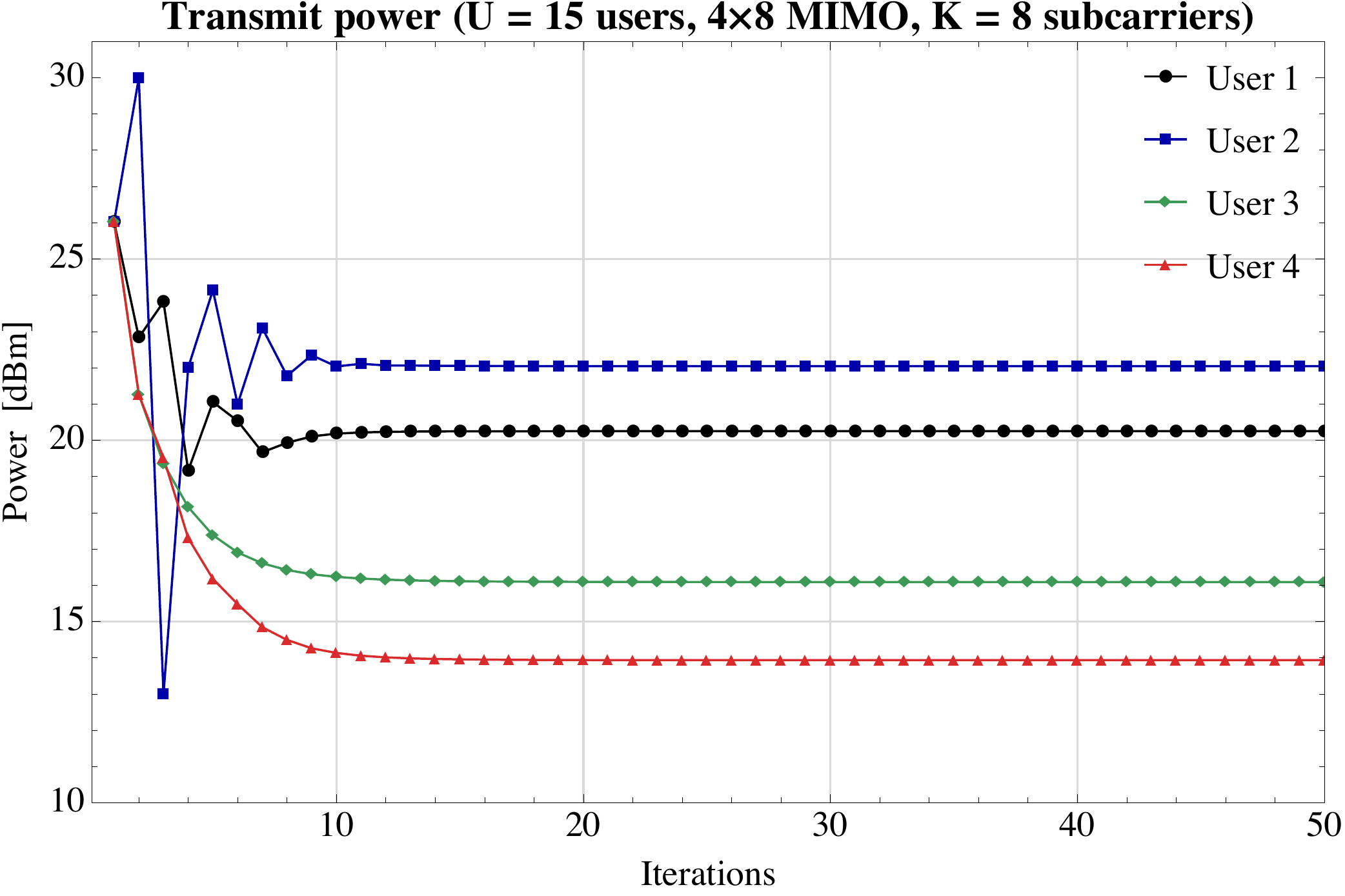}}
\hfill
\subfigure[Spectral efficiency evolution under \ac{OGA}]{\label{fig:static-rate}%
\includegraphics[width=.48\textwidth]{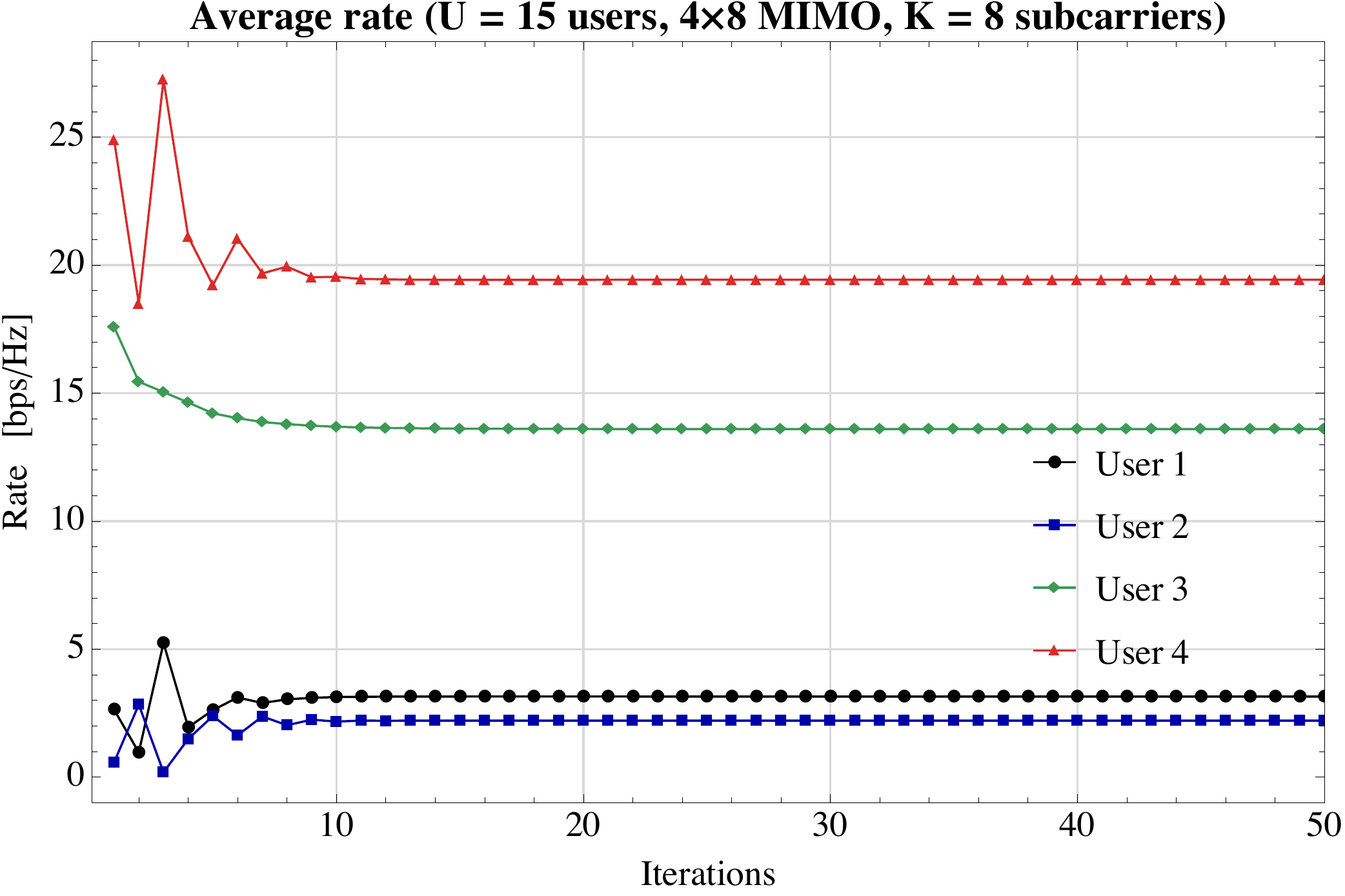}}
\\
\subfigure[Energy efficiency maximization under \ac{OGA}]{\label{fig:static-EE}%
\includegraphics[width=.48\textwidth]{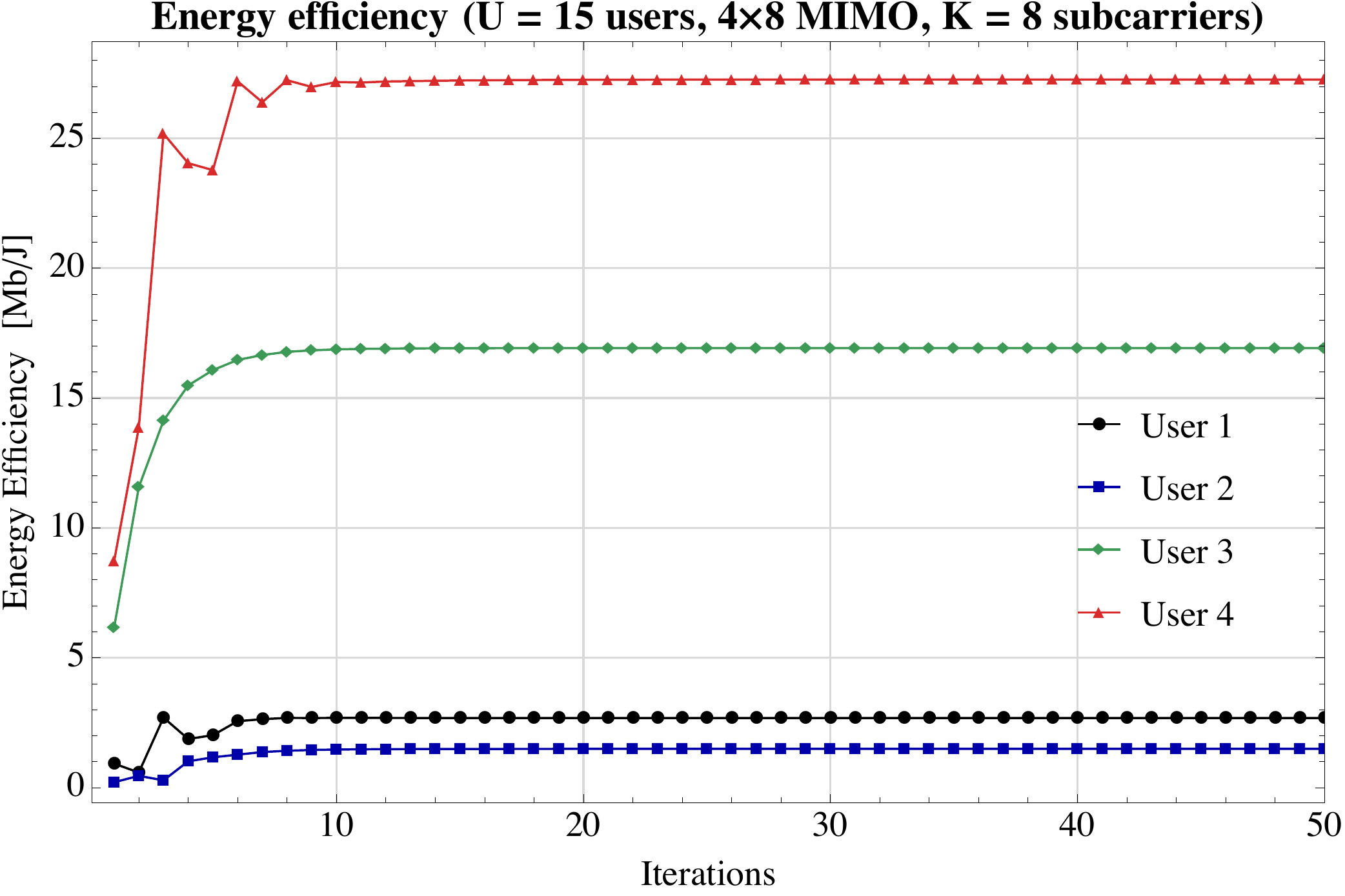}}
\hfill
\subfigure[Regret minimization under \ac{OGA}]{\label{fig:static-regret}%
\includegraphics[width=.48\textwidth]{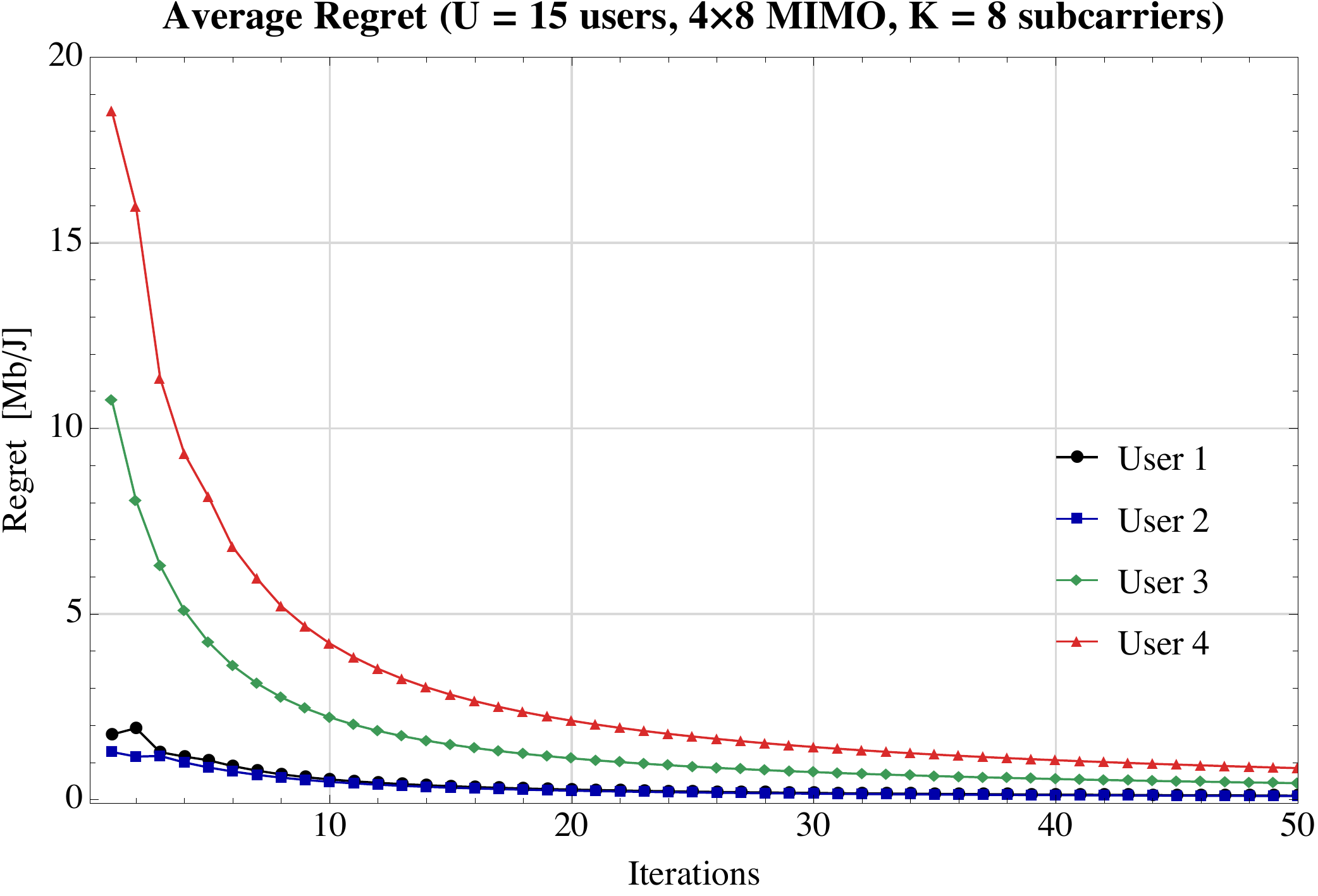}}
\\[-2ex]
\caption{Performance of the Algorithm~\ref{alg:OGA} under static channel conditions.
The system converges within a few iterations to an equilibrium state (Fig.~\ref{fig:static-EE}) where users experience no regret (Fig.~\ref{fig:static-regret}).}
\vspace{-2ex}
\label{fig:static}
\end{figure*}

\subsection{Static channels}
\label{sec:numerics-static}

For benchmarking purposes, our first simulation scenario addresses the case of stationary users with static channel conditions (so the variability of a user's effective channel matrix is only due to the modulation of the interfering users' transmit characteristics).
Each user is assumed to run \eqref{eq:OGA} with a variable step-size of the form $\step_{n} \propto 1/\sqrt{n}$
and an agnostic initialization with initial transmit power $P_{0} = \pmax/2 = 26\,\dbm$ spread evenly across antennas and subcarriers.
Our simulation results are presented in Fig.~\ref{fig:static} where, to minimize graphical clutter, we only plot the relevant information for $4$ users with diverse channel characteristics.

First, in Fig.~\ref{fig:static-power}, we plot the users' transmit power under \eqref{eq:OGA}.
As can be seen, even though users change their power by several \dbm, the algorithm quickly equilibrates after an initial transient phase.
Similarly, in Fig.~\ref{fig:static-rate}, we plot the users' transmit rate over all subcarriers (normalized by the bandwidth and thus measured in $\bps/\hz$).
We see here that users who reduce power by more than $10\,\dbm$ (Users $3$ and $4$) experience a commensurate drop in spectral efficiency (of the order of a few $\bps/\hz$);
on the other hand, users that decrease power only by a little achieve higher rates because the \ac{OGA} algorithm leads to a more efficient allocation of power over subcarriers and antennas.
Nonetheless, in all cases we observe a dramatic increase in \acl{EE} over the users' initial (uniform) power allocation policy, ranging from $\approx200\%$ to more than $600\%$.%
\footnote{Contrary to Fig.~\ref{fig:static-rate}, we do not normalize the users' \acl{EE} by the bandwidth, so it is measured in $\Mb/\joule$.}

In fact, as we see in Fig.~\ref{fig:static-EE}, after some slight oscillations during the first few iterations (the algorithm's transient phase), the system rapidly equilibrates and reaches a state where users have no incentive to change their individual power profiles (a \emph{Nash equilibrium}).
This equilibration is consistent with the no-regret properties of the \ac{OGA} algorithm:
as predicted by Theorem \ref{thm:reg-bound} and shown in Fig.~\ref{fig:static-regret}, the users' regret quickly decays to zero even though the algorithm's agnostic \textendash\ and, in hindsight, suboptimal \textendash\ initialization leads to high regret in the first few iterations.

\begin{figure*}[t!]
\footnotesize
\subfigure[Channel gain evolution for different user velocities]{\label{fig:varying-channels}%
\includegraphics[width=.4965\textwidth]{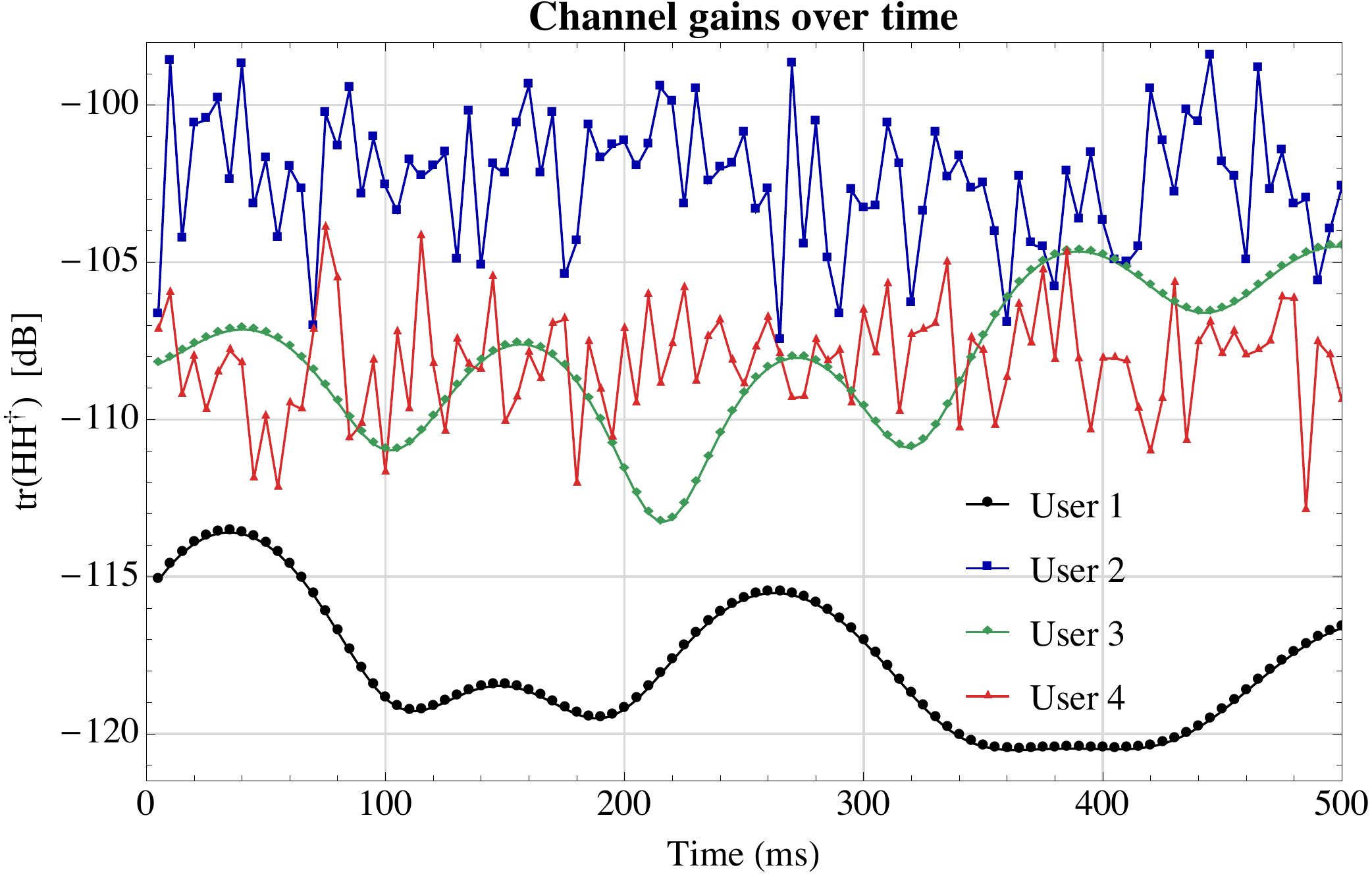}}
\hfill
\subfigure[User regret under \ac{OGA}]{\label{fig:varying-regret}%
\includegraphics[width=.48\textwidth]{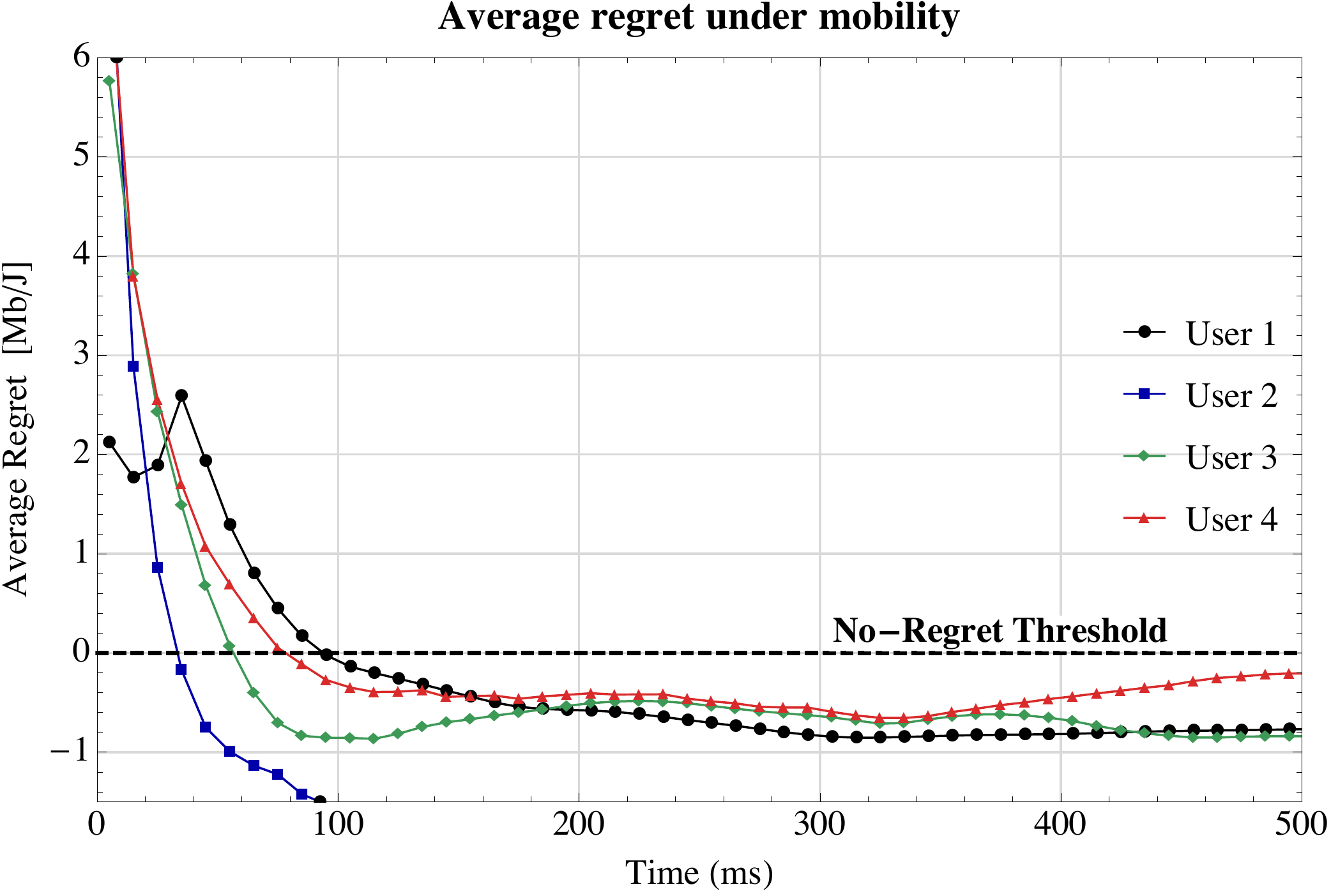}}
\\
\subfigure[Energy efficiency under \ac{OGA} (pedestrian)]{\label{fig:varying-EE-slow}%
\includegraphics[width=.48\textwidth]{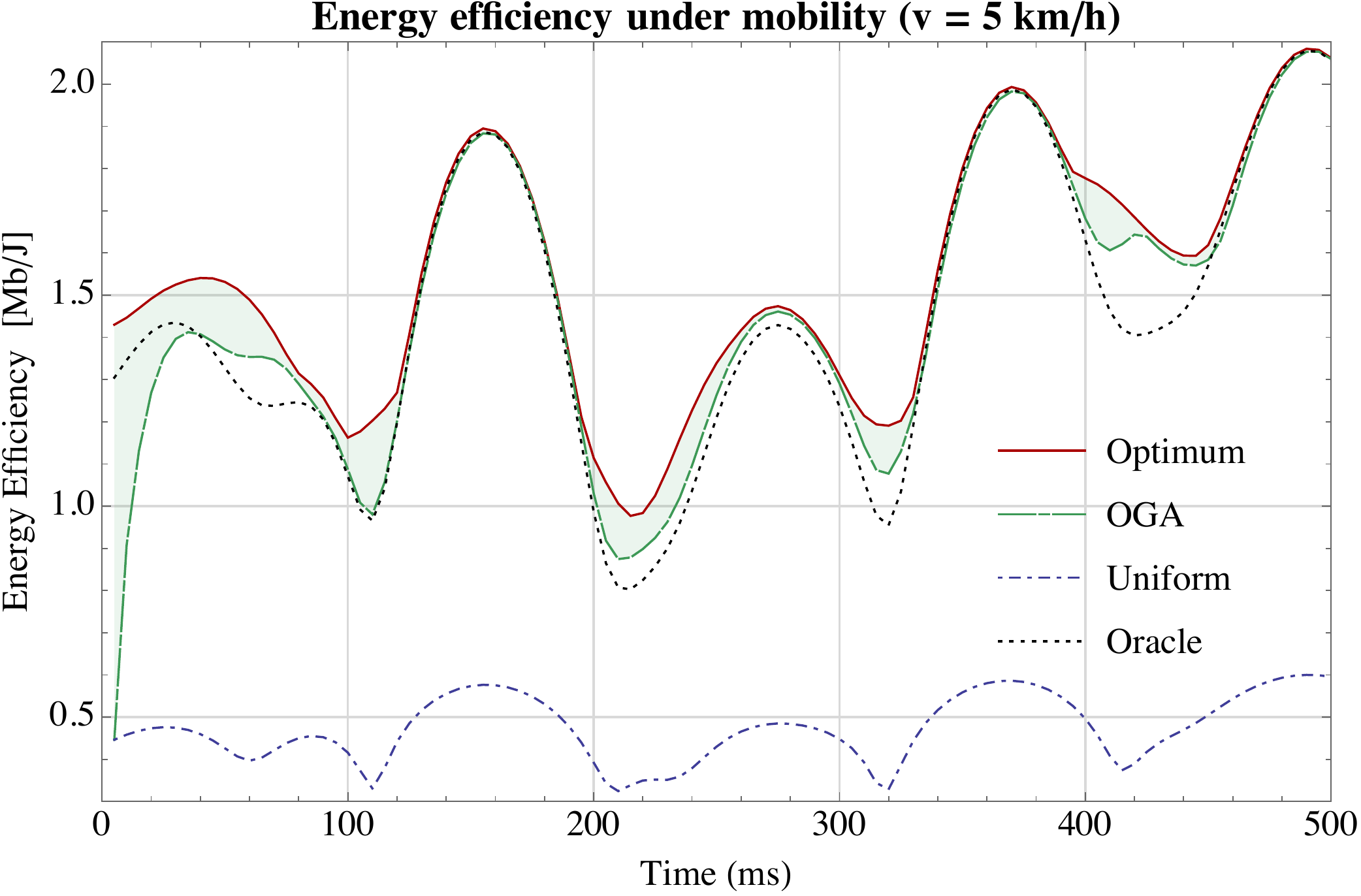}}
\hfill
\subfigure[Energy efficiency under \ac{OGA} (vehicular)]{\label{fig:varying-EE-fast}%
\includegraphics[width=.48\textwidth]{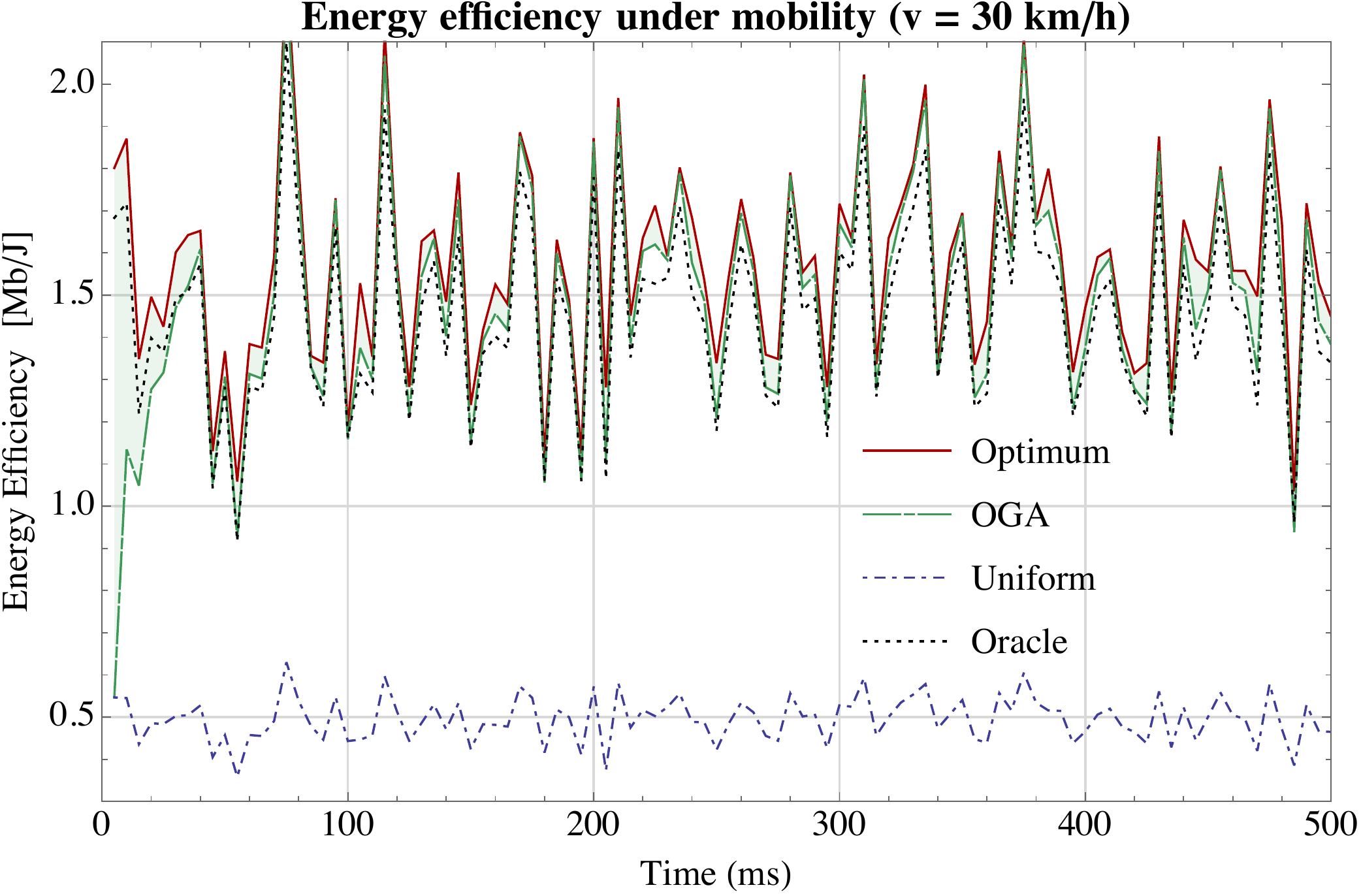}}
\\
\subfigure[Transmit power evolution under \ac{OGA} (pedestrian)]{\label{fig:varying-power-slow}%
\includegraphics[width=.48\textwidth]{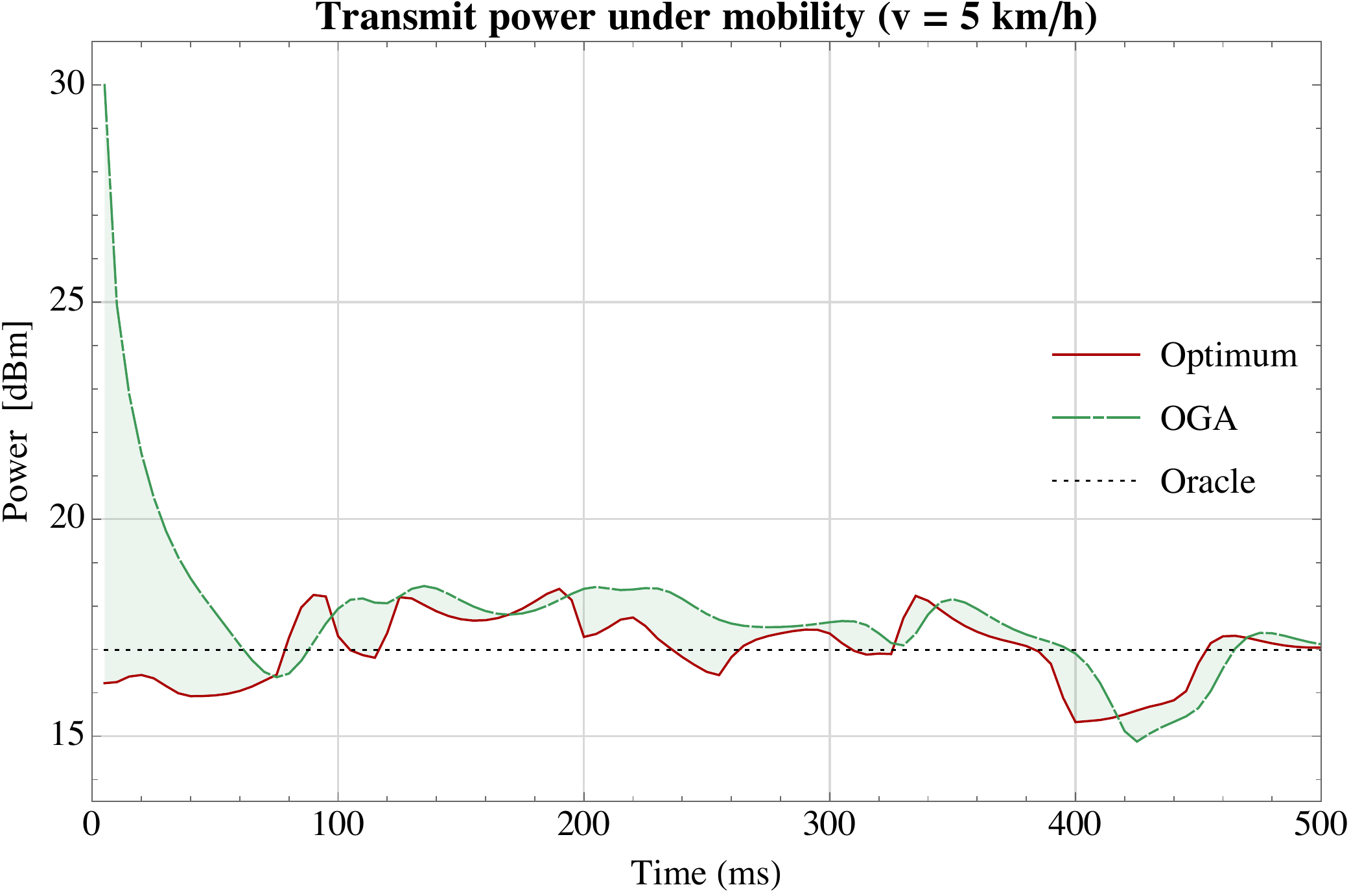}}
\hfill
\subfigure[Transmit power evolution under \ac{OGA} (vehicular)]{\label{fig:varying-power-fast}%
\includegraphics[width=.48\textwidth]{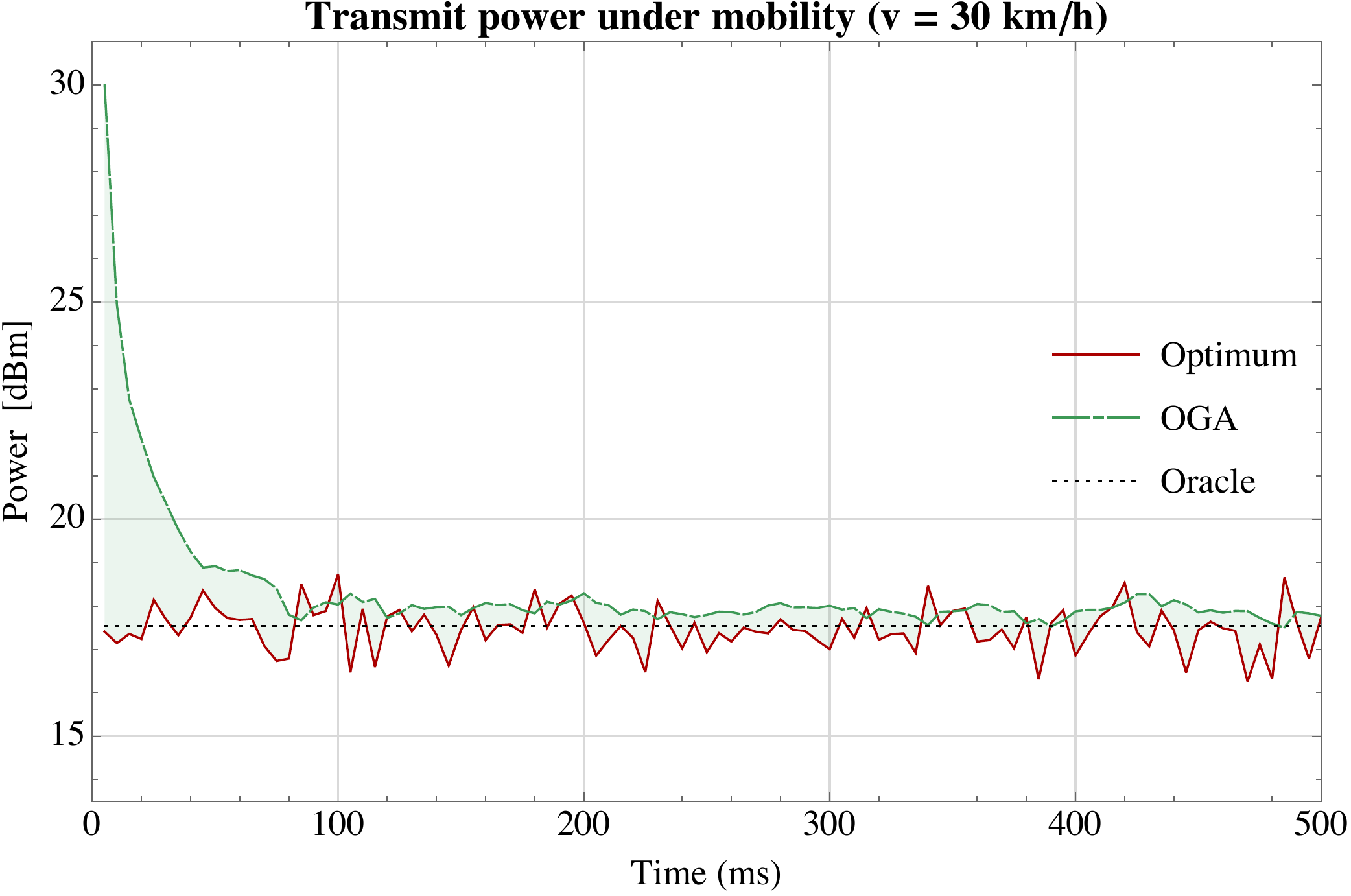}}
\\[-4ex]
\caption{Performance of the \ac{OGA} algorithm in a dynamic setting with mobile users moving at $v = \{3, 30, 5,130\}\,\kmh$.
The users' achieved \acl{EE} tracks its (evolving) maximum value remarkably well, even under rapidly changing channel conditions.}
\vspace{-2ex}
\label{fig:varying}
\end{figure*}

\subsection{Time-varying channels and mobility}
To account for dynamic network conditions, we also consider in Fig.~\ref{fig:varying} the case of mobile users whose channels vary with time due to Rayleigh fading, path loss fluctuations, etc.
For simulation purposes, we used the \ac{ETU} model for the users' environment and the \ac{EPA} and \ac{EVA} models to simulate pedestrian ($3$\textendash$5\,\kmh$) and vehicular ($30$\textendash$130\,\kmh$) movement respectively \cite{3GPP};
for reference, the focal users' channel gains ($\tr(\bH\bH^{\dag})$) have been plotted in Fig.~\ref{fig:varying-channels}.
Despite the channels' variability, Fig.~\ref{fig:varying-regret} shows that the users attain a no-regret state in a few iterations, even under rapidly changing channel conditions (cf. the case of Users $2$ and $4$ with an average speed of $30\,\kmh$ and $130\,\kmh$ respectively).
For completeness, we also plot in Figs.~\ref{fig:varying-EE-slow} and \ref{fig:varying-EE-fast} the achieved \acl{EE} for a pedestrian and a vehicular user, and we compare it to its instantaneous maximum value, the users' initial (uniform) power allocation policy, and the ``oracle'' solution which corresponds to the best fixed transmit profile in hindsight (i.e. the solution of the offline maximization problem \eqref{eq:EE-average} which posits that users can predict the system's evolution in advance).
Remarkably, even under rapidly changing channel conditions, the users' achieved \acl{EE} tracks its (evolving) maximum value remarkably well and consistently outperforms even the oracle solution (a fact which is consistent with the negative regret observed in Fig.~\ref{fig:varying-regret}).

An intuitive explanation for the adaptability of \ac{OGA} is provided by Figs.~\ref{fig:varying-power-slow} and \ref{fig:varying-power-fast} where we plot the transmit power of the optimum policy, the \ac{OGA} scheme, and the oracle solution for the same users as in Figs.~\ref{fig:varying-EE-slow} and \ref{fig:varying-EE-fast}.
Even though the optimum covariance matrix $\qtarget(n)$ may change significantly from one frame to the next, $\tr(\qtarget(n))$ remains roughly constant (within a few \dbm) over the entire transmission horizon.
The \ac{OGA} algorithm then learns this power level in a few iterations and stays close to it throughout the transmission horizon;
as a result, the users' achieved \acl{EE} remains itself very close to its maximum value for all time.

\begin{figure*}[t]
\includegraphics[width=.48\textwidth]{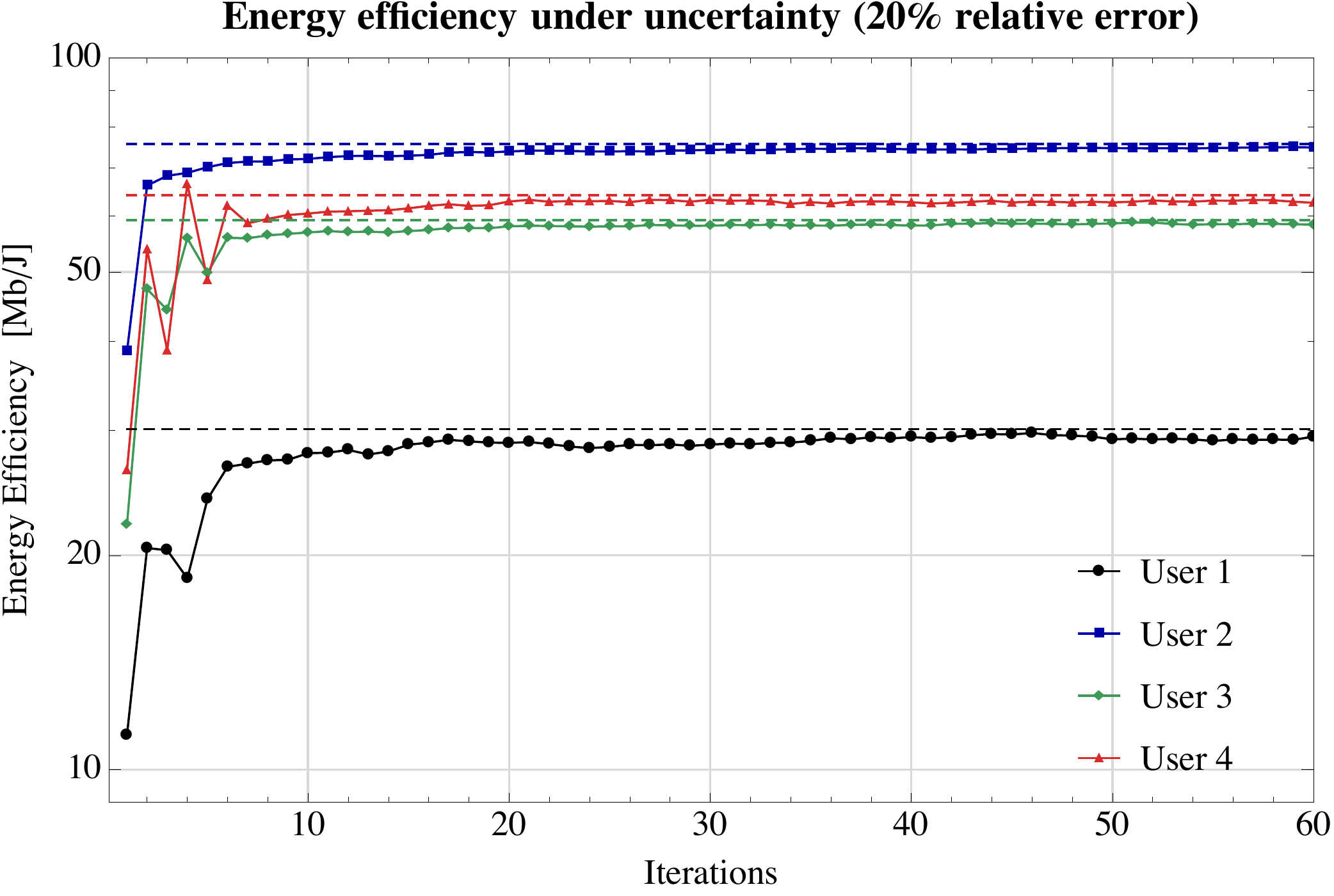}
\hfill
\includegraphics[width=.48\textwidth]{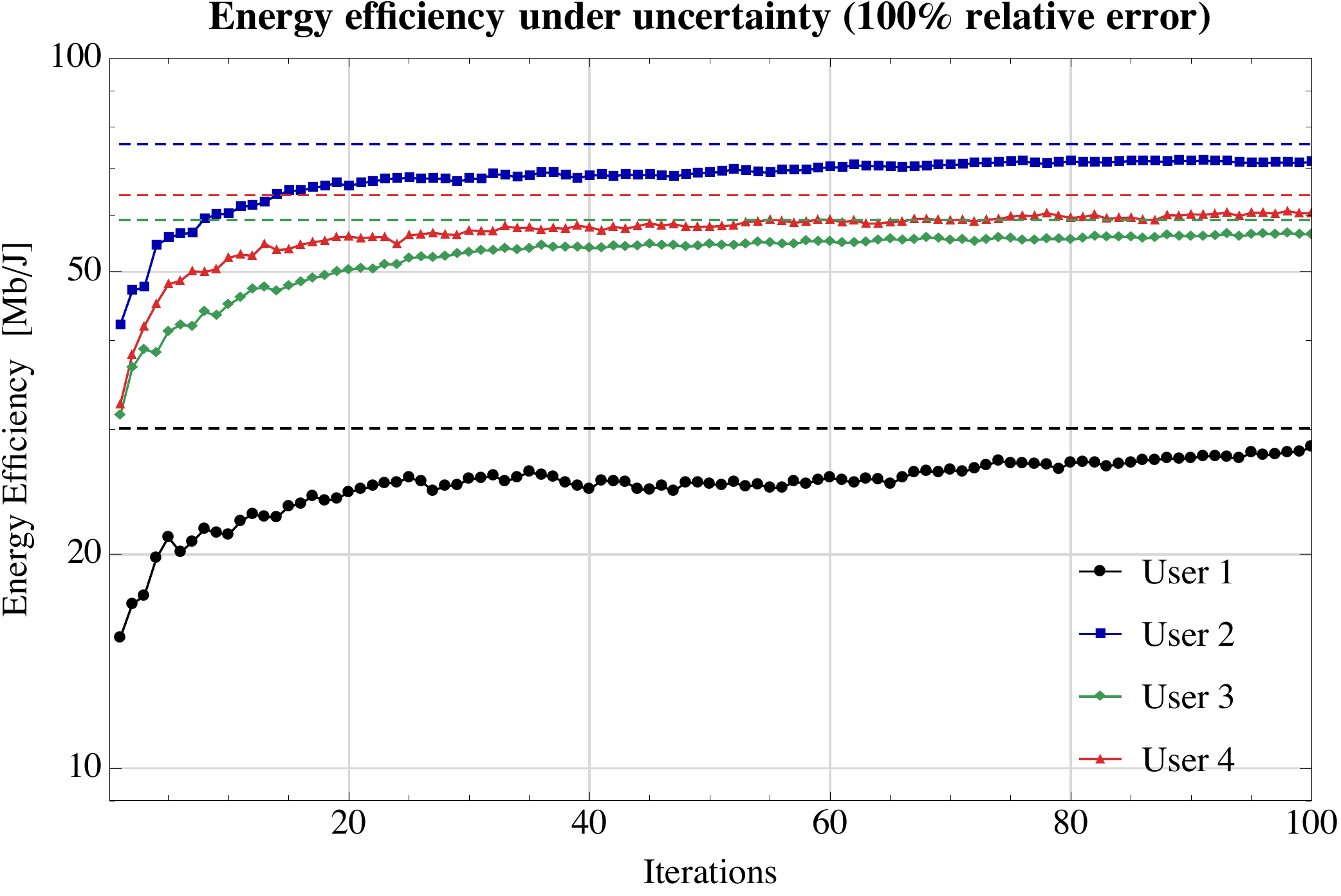}
\\[-3ex]
\caption{Performance of Algorithm~\ref{alg:OGA} with imperfect measurements and observation errors.
Remarkably, even under very high uncertainty, the system converges within a few tens of iterations to a stable state (dashed lines) where users experience no regret.}
\vspace{-3ex}
\label{fig:noisy}
\end{figure*}

\subsection{Robustness to observation noise and scalability for large antenna numbers}
\label{sec:numerics-noisy}

Finally, to assess the robustness of the \ac{OGA} algorithm in the presence of observation noise and measurement errors, the simulation cycle above was repeated in Fig.~\ref{fig:noisy} for the case where users only have access to noisy gradient observations as in Section \ref{sec:imperfectCSI}.
Also, to study the algorithm's scalability in the massive \ac{MIMO} regime (large number of antennas), we increased the number of transmit antennas to $\tx=8$ and the number of receive antennas to $\rx=128$;
otherwise, for comparison purposes, we used the same network simulation parameters as in Fig.~\ref{fig:static}.

The intensity of the measurement noise was quantified via the relative error level of the estimator $\hat\bV$, i.e. its standard deviation over its mean (so a relative error level of $\eta\%$ means that the observed matrix $\hat\bV$ lies within $\eta\%$ of its true value).
We then plotted the users' achieved \acl{EE} under the \ac{OGA} algorithm for noise levels $\eta = 20\%$ (moderate-to-high uncertainty) and $\eta = 100\%$ (very high uncertainty).
As can be seen in Fig.~\ref{fig:noisy}, the system's rate of equilibration is adversely affected by the intensity of the noise;
however, the system still equilibrates within a few tens of iterations and the users exhibit a drastic increase in \acl{EE} (of the order of $150\%$ and higher), even in the presence of very high measurement noise.

\section{Conclusions and Perspectives}
\label{sec:conclusions}

In the context of multi-carrier, massive \ac{MIMO} systems where numerous interfering mobile users co-exist, the temporal variability of the system cannot be ignored when targeting highly energy-efficient communications.
To tackle these issues, we introduced an online semidefinite optimization framework for the study of \acl{EE} maximization in dynamically varying networks, and we proposed an adaptive transmit policy that allows users to attain a ``no-regret'' state.
Importantly, the proposed policy is distributed, asynchronous, computationally simple, and it only requires minimal, strictly causal and (potentially) noisy \acl{CSIT}.
Specifically, under very mild assumptions for the statistics of the error process, we showed that the users' average regret after $\horizon$ epochs decays as $\bigoh(\horizon^{-1/2})$, a bound which is further improved to $\bigoh(\log \horizon/\horizon)$ if the users do not experience arbitrarily bad channel conditions.
As a result, users are able to track their most efficient transmit power profile with modest feedback requirements, even under rapidly changing channel conditions (corresponding to highly mobile users):
specifically, our simulations show that users could gain up to $600\%$ in \acl{EE} over fixed/uniform power allocation policies in realistic network environments.

An important theoretical question which arises is whether the system converges to an equilibrium state if all users employ a no-regret policy (our numerical simulations show that this indeed the case over a wide region of system parameters).
Additionally, different throughput-per-power models accounting for the probability of outage can also be considered and would require a modification of the proposed transmit policy.
We intend to explore these directions in future work.


\appendix[Technical Proofs]
\label{app:proofs}

Throughout this appendix (and unless explicitly mentioned otherwise), all matrices are assumed to be Hermitian and of dimension $D = K\tx$.
Additionally, the stage variable $n$ will be written as a subscript instead of as a parenthetical argument \textendash\ i.e. we will write $\bX_{n}$ and $\pay_{n}(\bX)$ instead of $\bX(n)$ and $\pay(\bX;n)$ respectively.
We do so in order to reduce the notational clutter caused by an overflow of parentheses;
since we will not require a subcarrier index, there is no fear of ambiguity.

\subsection{Matrix projections}
\label{app:proofs-projection}

We first prove that the projection map $\choice(\bY)$ is given by the explicit formula \eqref{eq:choice-explicit}.
To that end, simply note that $\choice(\bY)$ can be expressed equivalently as the solution to the maximization problem
\begin{equation}
\label{eq:proj-max}
\begin{aligned}
\text{maximize}
	&\quad
	\tr(\bY\bX) - \tfrac{1}{2} \norm{\bX}^{2},
	\\
\text{subject to}
	&\quad
	\bX\in\strat.
\end{aligned}
\end{equation}
However, if $\bY = \bU \bDelta \bU^{\dag}$ is a diagonalization of $\bY$,
the objective of \eqref{eq:proj-max} can be written as:
\begin{equation}
\tr(\bY\bX) - \tfrac{1}{2} \norm{\bX}^{2}
	= \tr(\bDelta \bU^{\dag} \bX \bU) - \tfrac{1}{2} \tr(\bU^{\dag}\bX\bU\bU^{\dag}\bX\bU).
\end{equation}
Thus, given that $\bX\in\strat$ if and only if $\bU\bX\bU^{\dag}\in\strat$, we readily get $\choice(\bY) = \bU \choice(\bDelta) \bU^{\dag}$, so it suffices to prove \eqref{eq:proj-max} for diagonal $\bY$.

We first show that $\choice(\bY)$ is itself diagonal if $\bY = \diag(\by)$ for some $\by\in\R^{D}$.
Indeed, we have:
\begin{equation}
\tr(\bY\bX) - \tfrac{1}{2} \norm{\bX}^{2}
	= \insum_{i} y_{i} X_{ii}^{2} - \tfrac{1}{2} \insum_{i,j} \smallabs{X_{ij}}^{2}
	\leq \insum_{i} \left( y_{i} X_{ii}^{2} - \tfrac{1}{2} X_{ii}^{2} \right),
\end{equation}
with equality if and only if $\bX$ is diagonal.
As a result, if $\bX$ is a solution of \eqref{eq:proj-max}, the diagonal matrix $\bX'$ which coincides with $\bX$ on the diagonal and has zero entries otherwise will also be a solution of \eqref{eq:proj-max};
since \eqref{eq:proj-max} admits a unique solution, we conclude that $\choice(\bY)$ must also be diagonal, as claimed.
We are thus left to solve the maximization problem
\begin{equation}
\label{eq:proj-simplex}
\begin{aligned}
\text{maximize}
	&\quad\txs
	\sum_{j} y_{j} x_{j} - \tfrac{1}{2} \sum_{j} x_{j}^{2},
	\\
\text{subject to}
	&\quad\txs
	x_{j} \geq 0,\;\;
	\sum_{j} x_{j} \leq 1.
\end{aligned}
\end{equation}

Writing $\lambda_{j} \geq 0$ and $\lambda \geq 0$ for the Lagrange multipliers of the constraints $x_{j} \geq 0$ and $\sum_{j} x_{j} \leq 1$ respectively, the first-order \ac{KKT} conditions for \eqref{eq:proj-simplex} become:
\begin{subequations}
\label{eq:KKT}
\begin{flalign}
\label{eq:KKT-obj}
&\txs
y_{j} = x_{j} + \lambda - \lambda_{j},
	\\
\label{eq:KKT-ineq}
&\txs
\lambda_{j} x_{j} = 0,\;\;
	\lambda \,\big(1 - \sum_{j} x_{j} \big) = 0.
\end{flalign}
\end{subequations}
Thus, to obtain the first branch of \eqref{eq:smallchoice}, simply note that if $y_{j}\leq0$ but $x_{j}>0$, we will also have $\lambda_{j} = 0$, so \eqref{eq:KKT-obj} gives $y_{j} = x_{j} + \lambda > 0$, a contradiction.
Likewise, if $\sum_{j: y_{j} \geq 0} y_{j} \leq 1$, setting $\lambda_{j} = \lambda = 0$ and $x_{j} = y_{j}$ for all $j$ such that $y_{j} \geq 0$ is obviously a solution of \eqref{eq:KKT}, so we obtain the second branch of \eqref{eq:smallchoice}.
Finally, to obtain the third branch of \eqref{eq:smallchoice}, note first that $\sum_{j:y_{j}\geq0} x_{j} = 1$ if $\sum_{j:y_{j}\geq0} y_{j} \geq 1$;
otherwise, we would have $\lambda=0$ and \eqref{eq:KKT-obj} would give $y_{j} = x_{j} - \lambda_{j} \leq x_{j}$ whenever $y_{j}\geq0$, implying in turn that $\sum_{j:y_{j}\geq0} y_{j} \leq \sum_{j:y_{j}\geq0} x_{j} < 1$, a contradiction.
Accordingly, we are left to project the vector with components $y_{i}^{+} = \pospart{y_{i}}$ to the unit simplex $\simplex = \{\bx\in\R^{D}: x_{j}\geq0 \text{ and } \sum_{j} x_{j} = 1\}$;
this projection simply gives $x_{i} = [y_{i}^{+}-\lambda]_{+}$ with $\lambda \geq 0$ such that $\sum_{i} [y_{i}^{+} - \lambda]_{+} = 1$ \cite{MdP89}, so \eqref{eq:choice-explicit} follows.



\subsection{No regret with perfect \ac{CSI}}
\label{app:proofs-reg-bound}

The key step in bounding the user's regret is the inequality:
\begin{equation}
\label{eq:reg-gradient}
\ee(\qtarget) - \ee(\bQ)
	= \pay(\target) - \pay(\bX)
	\leq \tr[\bV\cdot (\target - \bX)],
\end{equation}
which is a simple consequence of the fact that $\pay(\bX)$ is concave in $\bX$.
Our proof follows the methodology of \cite{Zin03} where \ac{OGA} methods where used in a vector (as opposed to matrix) setting with the special step-size sequence $\step_{n} \propto n^{-1/2}$ (as opposed to general $\step_{n}$).
To be precise, we will establish the no-regret properties of \eqref{eq:OGA} by showing that $\sum_{n=1}^{\horizon} \tr[\bV_{n}\cdot(\target - \bX_{n})] = o(\horizon)$ for all $\target\in\strat$ and for every matrix sequence $\bV_{n}$.

\begin{IEEEproof}[Proof of Theorem \ref{thm:reg-bound}]
Letting $\breg_{n} = \frac{1}{2} \norm{\target - \bX_{n}}^{2}$, we get:
\begin{equation}
\label{eq:Breg1}
D_{n+1}
	= \frac{1}{2} \norm{\target - \bX_{n+1}}^{2}
	= \frac{1}{2} \norm{\target - \choice(\bX_{n} + \step_{n} \bV_{n})}^{2}
	\leq \frac{1}{2} \norm{\target - \bX_{n} - \step_{n} \bV_{n}}^{2},
\end{equation}
on account of the definition of $\choice(\bY)$ as the closest point to $\bY$ on $\strat$.
In this way, \eqref{eq:Breg1} yields:
\begin{equation}
\label{eq:Breg2}
\breg_{n+1}
	\leq \breg_{n} - \step_{n} \tr\left[ \bV_{n}\cdot(\target - \bX_{n}) \right] + \frac{1}{2} \step_{n}^{2} \norm{\bV_{n}}^{2},
\end{equation}
and hence, after rearranging and summing over $n$, we obtain:
\begin{flalign}
\label{eq:Breg3}
\sum_{n=1}^{\horizon} \tr\left[ \bV_{n} \cdot (\target - \bX_{n}) \right]
	&\leq \sum_{n=1}^{\horizon} \step_{n}^{-1} \left( \breg_{n} - \breg_{n+1} \right)
	+ \frac{1}{2} \sum_{n=1}^{\horizon} \step_{n} \norm{\bV_{n}}^{2}
	\notag\\
	&\leq \step_{1}^{-1}  \breg_{1}
	+ \sum_{n=2}^{\horizon} \left( \step_{n}^{-1} - \step_{n-1}^{-1} \right) \breg_{n}
	+ \frac{1}{2} \sum_{n=1}^{\horizon} \step_{n} \norm{\bV_{n}}^{2}
	\notag\\
	&\leq \step_{1}^{-1} \depth + \sum_{n=2}^{\horizon} \left( \step_{n}^{-1} - \step_{n-1}^{-1} \right) \depth
	+ \frac{\vbound^{2}}{2} \sum_{n=1}^{\horizon} \step_{n}
	= \frac{1}{\step_{\horizon}} + \frac{\vbound^{2}}{2} \sum_{n=1}^{\horizon} \step_{n},
\end{flalign}
where $\depth \equiv \frac{1}{2} \max_{\bX,\bX'\in\strat} \norm{\bX - \bX'}^{2} = 1$.
The fact that $\bX_{n}$ leads to no regret then follows by noting that $1/(\horizon \step_{\horizon}) \to 0$ (by assumption) and that $\horizon^{-1} \sum_{n=1}^{\horizon} \step_{n} \to 0$ (since $\step_{n}\to 0$).
\end{IEEEproof}

\begin{IEEEproof}[Proof of Proposition \ref{prop:reg-log}]
We note first that the $a$-strong concavity assumption \eqref{eq:pay-strong} for $\pay$ gives:
\begin{equation}
\label{eq:pay-Taylor}
\pay_{n}(\target) - \pay_{n}(\bX_{n})
	\leq \tr\left[ \bV_{n} (\target - \bX_{n}) \right] - \frac{1}{2\step} \norm{\target - \bX_{n}}^{2},
\end{equation}
where we have used the fact that $a \geq \step^{-1}$.
Thus, by summing over $n$ and using \eqref{eq:Breg3}, we obtain:
\begin{flalign}
\label{eq:reg-step-1}
\reg(\horizon)
	&\leq \frac{1}{2} \sum_{n=2}^{\horizon} \left( \step_{n}^{-1} - \step_{n-1}^{-1} - \step^{-1} \right) \norm{\target - \bX_{n}}^{2}
	+ \frac{1}{2} \sum_{n=1}^{\horizon} \step_{n} \norm{\bV_{n}}^{2}
	\notag\\
	& \leq \frac{1}{2} \step \vbound^{2} \sum_{n=1}^{\horizon} n^{-1}
	\leq \frac{1}{2} \step \vbound^{2} (1 + \log\horizon),
\end{flalign}
where, in the second line, we used the fact that $\step_{n}^{-1} - \step_{n-1}^{-1} = n\step^{-1} - (n-1) \step^{-1} = \step^{-1}$.
\end{IEEEproof}

\subsection{The case of imperfect \ac{CSI}}
\label{app:proofs-reg-stoch}

\begin{IEEEproof}[Proof of Theorem \ref{thm:reg-stoch}]
As before, we begin with the basic inequality:
\begin{equation}
\label{eq:reg-stoch0}
\reg(\horizon)
	= \max_{\target\in\strat} \insum_{n=1}^{\horizon}
	\left[ \pay_{n}(\target) - \pay_{n}(\bX_{n}) \right]
	\leq \max_{\target\in\strat} \insum_{n=1}^{\horizon}
	\tr\left[ \bV_{n}\cdot (\target - \bX_{n}) \right],
\end{equation}
with $\bX_{n}$ defined via the (stochastic) recursion:
\begin{equation}
\label{eq:OGA-stoch}
\bX_{n+1}
	= \choice(\bX_{n} + \step_{n} \hat\bV_{n}).
\end{equation}
Thus, for the first part of the theorem, we need to show that:
\begin{equation}
\label{eq:reg-stoch1}
\sum_{n=1}^{\horizon} \tr\big[ \hat\bV_{n}\cdot (\target - \bX_{n}) \big]
	+ \sum_{n=1}^{\horizon} \tr\left[ \bZ_{n}\cdot (\bX_{n} - \target) \right]
	= o(\horizon)
	\quad
	\text{(a.s.)},
\end{equation}
where $\bZ_{n} = \hat\bV_{n} - \bV_{n}$.
The first term of \eqref{eq:reg-stoch1} then becomes:
\begin{flalign}
\label{eq:reg-stoch2}
\sum_{n=1}^{\horizon} \tr\big[ \hat\bV_{n}\cdot (\target - \bX_{n}) \big]
	&\leq \step_{1}^{-1} \depth
	+ \sum_{n=2}^{\horizon} \left( \step_{n}^{-1} - \step_{n-1}^{-1} \right) \breg_{n}
	+ \frac{1}{2} \sum_{n=1}^{\horizon} \step_{n} \smallnorm{\hat\bV_{n}}^{2}
	\notag\\
	&\leq \step_{1}^{-1} \depth
	+ \sum_{n=2}^{\horizon} \left( \step_{n}^{-1} - \step_{n-1}^{-1} \right) \depth
	+ \frac{1}{2} \sum_{n=1}^{\horizon} \step_{n} \left(
	\norm{\bV_{n}}^{2} + 2 \tr(\bV_{n}\bZ_{n}) + \norm{\bZ_{n}}^{2}
	\right)
	\notag\\
	&= \frac{1}{\step_{\horizon}}
	+ \frac{\vbound^{2}}{2} \sum_{n=1}^{\horizon} \step_{n}
	+ \bigoh\left( \sum_{n=1}^{\horizon} \step_{n} \norm{\bZ_{n}}^{2} \right),
\end{flalign}
where, as before, $\depth \equiv \frac{1}{2} \max_{\bX,\bX'\in\strat} \norm{\bX - \bX'}^{2} = 1$.
We now claim that $\lim_{\horizon\to\infty} \horizon^{-1} \sum_{n=1}^{\horizon} \step_{n} \norm{\bZ_{n}}^{2}\to 0$ (a.s.).
Indeed, let $z_{n} = \smallnorm{\bZ_{n}}$ and choose $\eps >0$ such that $4\eps \leq \alpha\beta - 2$ (recall that $\alpha\beta>2$);
Hypothesis \eqref{eq:tailbound} then implies that $\prob\big(z_{n} \geq n^{\alpha/2 - \eps/\beta} \big) \leq B/n^{\alpha\beta/2 - \eps} \leq B/n^{1+\eps}$ for all $n$, so we obtain:
\begin{equation}
\sum_{n=1}^{\infty} \prob\left(z_{n} \geq n^{\alpha/2 - \eps/\beta} \right)
	= \sum_{n=1}^{\infty} \bigoh(1/n^{1+\eps})
	< \infty,
\end{equation}
and hence, by the Borel-Cantelli lemma, we conclude that
\(
\prob\big(
	\text{$z_{n} \geq n^{\alpha/2 - \eps/\beta}$ for infinitely many $n$}
	\big)
	= 0.
\)
In turn, this implies that $z_{n}^{2} = \bigoh\big(n^{\alpha - 2\eps/\beta} \big)$ almost surely, so we get:
\begin{equation}
\label{eq:reg-stoch3}
\sum_{n=1}^{\horizon} \step_{n} \norm{\bZ_{n}}^{2}
	= \bigoh\left( \sum_{n=1}^{\horizon} n^{-\alpha} n^{\alpha - 2\eps/\beta} \right)
	= \bigoh\left( \sum_{n=1}^{\horizon} 1/n^{2\eps/\beta} \right)
	= o(\horizon)
	\quad
	\text{(a.s.)}.
\end{equation}

For the second term of \eqref{eq:reg-stoch1}, let $\xi_{n} = \tr\left[ \bZ_{n}\cdot(\target - \bX_{n}) \right]$.
Then, given that $\bX_{n}$ is a deterministic function of $\bX_{n-1}$ and $\hat\bV_{n-1}$, we will also have $\ex\left[\xi_{n} \given \bX_{n-1} \right] = 0$, i.e. $\xi_{n}$ is a sequence of martingale differences.
By the strong law of large numbers for martingale differences \cite[Theorem~2.18]{HH80}, it then follows that $\lim_{\horizon\to\infty} \horizon^{-1} \sum_{n=1}^{\horizon} \xi_{n} = 0$ (a.s.).
As a result, combining this with \eqref{eq:reg-stoch3}, we get
\begin{equation}
\sum_{n=1}^{\horizon} \tr\big[ \hat\bV_{n} \cdot (\target - \bX_{n}) \big]
	= o(\horizon)
	\quad
	\text{(a.s.),}
\end{equation}
i.e. \eqref{eq:OGA-stoch} leads to no regret, as claimed.

Finally, for the mean regret bound \eqref{eq:reg-mean}, taking the expectation of the first line of \eqref{eq:reg-stoch2} yields:
\begin{flalign}
\label{eq:reg-stoch4}
\ex[\reg(\horizon)]
	&\leq \step_{1}^{-1}\depth
	+ \sum_{n=2}^{\horizon} \left( \step_{n}^{-1} - \step_{n-1}^{-1} \right) \breg_{n}
	+ \frac{1}{2} \sum_{n=1}^{\horizon} \step_{n} \ex\big[ \smallnorm{\hat\bV_{n}}^{2} \big]
	\leq \frac{1}{\step_{\horizon}} + \frac{\stochbound^{2}}{2} \sum_{n=1}^{\horizon} \step_{n},
\end{flalign}
where we used the fact that $\reg(\horizon) \leq \ex[\bV_{n}\cdot(\target - \bX_{n})] = \ex[\hat\bV_{n}\cdot(\target - \bX_{n})]$ for the LHS, and the assumption that $\ex\big[ \smallnorm{\hat\bV_{n}}^{2} \big] \leq \stochbound^{2}$ for the RHS.
\end{IEEEproof}

\begin{IEEEproof}[Proof of Proposition \ref{prop:reg-log-stoch}]
By reasoning as in the proof of Proposition \ref{prop:reg-log}, we readily obtain:
\begin{equation}
\label{eq:reg-log-stoch1}
\sum_{n=1}^{\horizon} \tr\big[ \hat\bV_{n} \cdot (\target - \bX_{n}) \big]
	\leq \frac{1}{2} \sum_{n=1}^{\horizon} \step_{n} \smallnorm{\hat\bV_{n}}^{2},
\end{equation}
so \eqref{eq:reg-log-mean} follows by taking expectations on both sides as in the proof of Theorem \ref{thm:reg-stoch}.
That $\bX_{n}$ leads to no regret then follows by noting that \eqref{eq:reg-stoch3} holds even for $\alpha=1$, so the RHS of \eqref{eq:reg-log-stoch1} is $o(\horizon)$.
\end{IEEEproof}

\bibliographystyle{IEEEtran}
\footnotesize
\setlength{\bibsep}{0pt}
\bibliography{IEEEabrv,EnergyEfficiency,VPbiblio}

\end{document}